\documentclass[prd,twocolumn,amsmath,amssymb,floatfix, superscriptaddress,nofootinbib]{revtex4}

\usepackage{bm}
\usepackage{amsmath}
\usepackage{epsf}
\usepackage{color}
\usepackage{natbib}
\usepackage{graphicx}

\newcommand{\sun}{\odot}

\newcommand{\eqn}[1]{Eq.~(\ref{eq:#1})}

\newcommand{\om}{\Omega_{\rm m}}
\newcommand{\omhh}{\Omega_{\rm m}h^2}

\newcommand{\ok}{\Omega_{\rm K}}

\newcommand{\ode}{\Omega_{\rm DE}}
\newcommand{\orad}{\Omega_{\rm r}}
\newcommand{\olam}{\Omega_{\Lambda}}

\newcommand{\zmax}{z_{\rm max}}
\newcommand{\emax}{E_{\rm max}}

\newcommand{\zminsn}{z_{\rm min}^{\rm SN}}
\newcommand{\nzpc}{N_{z,{\rm PC}}}

\newcommand{\wmin}{w_{\rm min}}
\newcommand{\wmax}{w_{\rm max}}
\newcommand{\wfid}{w_{\rm fid}}
\newcommand{\lcdm}{$\Lambda$CDM}
\newcommand{\winf}{w_{\infty}}
\newcommand{\scrm}{\mathcal{M}}

\newcommand{\rhode}{\rho_{\rm DE}}
\newcommand{\rhoc}{\rho_{{\rm cr},0}}
\newcommand{\dlum}{d_{\rm L}}

\newcommand{\tchi}{D_f}
\newcommand{\fcurv}{\kappa}
\newcommand{\cede}{C_{\rm EDE}}
\newcommand{\ccurv}{C_{\rm curv}}

\newcommand{\dpmax}{\delta'_{\rm max}}
\newcommand{\erf}{{\rm erf}}

\begin{document}

\title{Testing flatness of the universe with probes of cosmic distances and growth}

\author{Michael J. Mortonson}\email{mjmort@uchicago.edu}
\affiliation{Department of Physics, University of Chicago, Chicago IL 60637}
\affiliation{Kavli Institute for Cosmological Physics and Enrico Fermi Institute, University of Chicago, Chicago IL 60637, U.S.A.}

\date{\today}

\begin{abstract}
When using distance measurements to probe spatial curvature, 
the geometric degeneracy between curvature and dark energy 
in the distance--redshift relation 
typically requires either making 
strong assumptions about the dark energy evolution or sacrificing 
precision in a more model-independent approach.
Measurements of the redshift evolution of the linear growth of 
perturbations can break the geometric degeneracy, providing 
curvature constraints that are both precise and model-independent.
Future supernova, CMB, and cluster data have the potential 
to measure the curvature 
with an accuracy of $\sigma(\ok)=0.002$, 
without specifying a particular dark energy phenomenology.
In combination with distance measurements, 
the evolution of the growth function at low redshifts provides the strongest 
curvature constraint if the high-redshift universe is well 
approximated as being purely matter dominated.
However, in the presence of early dark energy or massive neutrinos,
the precision in curvature is reduced due to additional degeneracies, and
precise normalization of the growth function relative to 
recombination is important for obtaining accurate constraints.
Curvature limits from distances and growth compare favorably to 
other approaches to curvature
estimation proposed in the literature, providing either greater
accuracy or greater freedom from dark energy modeling assumptions, 
and are complementary due to the use of independent data sets.
Model-independent estimates of curvature are 
critical for both testing inflation and obtaining unbiased constraints 
on dark energy parameters.
\end{abstract}

\maketitle

\defcitealias{Mortonson09a}{MHH09}

\section{Introduction}
\label{sec:intro}

Cosmological measurements of the average spatial curvature 
of the spacetime metric are 
one of only a handful of methods available for testing the 
inflationary paradigm for the early universe. 
Current observations are consistent with the inflationary 
prediction of a nearly flat universe.
However, the precision of curvature measurements is only at the 
percent level at best (e.g., \cite{Eisenstein05a,Komatsu09a}), 
whereas the expected level of curvature in standard inflationary scenarios
is generally much smaller. 
Moreover, obtaining the most precise limits on curvature requires assuming 
a particular simple form for the dark energy evolution due 
to the well-known ``geometric degeneracy'' between curvature 
and dark energy \citep{Weinberg70a,Bond97a,Zaldarriaga97a,Efstathiou99a,Huey99a,Caldwell04a,Linder05a,Hu05a,Polarski05a,Bernstein06a,Knox06a,Hu06b,Clarkson07a,Zhao07a,Ichikawa07a,Wright07a,Huang07a,Hlozek08a,Virey08a}.
Inferences about inflation based on such curvature constraints are 
only valid if the assumed dark energy behavior is an 
adequate description of the true evolution.

Just as uncertainty about the dark energy evolution affects estimates 
of curvature, uncertainty about curvature limits our ability to 
constrain parameters of dark energy models with cosmic distances.
One often assumes spatial flatness motivated by the predictions of inflation when 
constraining dark energy models, but the resulting parameter estimates may 
be biased if the true spatial curvature deviates even slightly from zero
\citep{Linder05a,Clarkson07a,Zhao07a,Hlozek08a,Virey08a}.

A few methods for using measured distances to obtain curvature estimates 
that are \emph{independent} of the dark energy evolution
have been proposed for use with future data sets.
For example, Bernstein \cite{Bernstein06a} proposed a 
technique using weak lensing galaxy--shear correlations to measure 
triangles of distances between the lensing and source galaxy planes 
and the observer, resulting in a curvature estimate that depends only 
on the assumed form of the spacetime metric.
An alternate method by Knox \cite{Knox06a} uses precisely measured 
distances at high redshift ($z\gtrsim 3$) combined with the distance 
to recombination from cosmic microwave background (CMB) data to infer the curvature 
without dependence on the low-redshift dark energy evolution.

In this work, we describe a different approach to model-independent curvature estimates that
uses combinations of data sets that probe the distance--redshift relation 
and the growth of linear perturbations. The geometric degeneracy in 
distance data arises because distances depend on both the expansion rate 
(and therefore the dark energy evolution) and the spatial curvature.
The growth of structure, on the other hand, depends only on the 
expansion rate.  If one can measure both distances and growth over a similar 
range of redshifts, then constraints on the expansion rate from 
growth data can be used to break the degeneracy in distance data, 
providing a model-independent determination of the curvature.
The only assumptions required are 
that general relativity (GR) is the correct theory of 
gravity governing the growth of structure and that dark energy does 
not cluster significantly on the scales of interest.

In Section~\ref{sec:pre}, we review the 
basic distance and growth relations and observables.
Section~\ref{sec:degen} describes the geometric degeneracy in distances 
and how this degeneracy is broken by growth information.
We then present forecasts for curvature 
constraints from future distance and growth data, 
beginning with descriptions of two methods of relating growth evolution 
on linear scales to an observed distance--redshift relation.
The first, in Sec.~\ref{sec:mcmc}, is based on a 
numerical exploration of general dark energy models 
carried out in Ref.~\cite{Mortonson09a} by Mortonson, Hu, and Huterer 
(hereafter, MHH09). 
The second method involves rewriting the equations
for distances and growth so that the common dependence on the 
expansion rate drops out.
The basis of this method of reconstructing the growth history from 
observed distances comes from Alam, Sahni, and Starobinsky \cite{Alam08a}, 
and in Sec.~\ref{sec:analytic}, we summarize this work 
and extend it in several ways to allow the method to be applied to 
curvature forecasts. These two methods are complementary in several ways;
the MCMC approach is more straightforward in terms of 
error propagation and the solution for the growth evolution, 
but it can be quite time-consuming and depends more on one's priors
on the dark energy evolution compared with the analytic 
growth reconstruction approach.
The growth reconstruction method is therefore a useful tool for
exploring the curvature-dependent relation between distances and growth 
for a variety of different cosmologies and assumed data sets, 
while the MCMC results help in testing and calibrating the analytic method 
and in providing accurate error estimates.
Using both of these methods, forecasts for curvature from a combination of 
future supernova (SN), 
CMB, and X-ray cluster data are 
presented in Sec.~\ref{sec:curv}.
Finally, Sec.~\ref{sec:conc} 
contains a summary and discussion of the results of this work.

\vspace{.5cm}

%%%%%%%%%%%%%%%%%%%%%%%%%%%%%%%%%%%%%%%%%%%%%%%%%%%%%%%%%%%%%%%%%
\section{Preliminaries}
\label{sec:pre}

%########################################
\subsection{Spatial curvature}
\label{sec:curvintro}

Given that the universe appears to be 
spatially homogeneous and isotropic on large scales, the background metric 
can be written in the Friedmann-Robertson-Walker (FRW) form:
\begin{equation}
ds^2 = -dt^2 + a^2\left[\frac{dD^2}{1+\ok H_0^2 D^2}
 +D^2(d\theta^2 + \sin^2 \theta~d\phi^2)\right],
\label{eq:frw}
\end{equation}
which describes an expanding (or contracting) universe with 
scale factor $a(t)$, where $D$ is the comoving radial coordinate 
and $H_0$ is the Hubble constant.
The FRW metric has constant spatial curvature 
parametrized by $\ok$, where a flat universe has $\ok=0$, 
an open universe $\ok>0$, and a closed universe $\ok<0$.
The curvature parameter is related to the 
total density of the components of the universe in units of 
the critical density for flatness, 
$\Omega_{\rm tot}=\rho_{\rm tot}/\rhoc$, by 
$\ok=1-\Omega_{\rm tot}$, where all densities are evaluated 
at the present time.

Theories of inflation predict that the universe 
is nearly flat ($\ok\approx 0$), and the fact that current observations
are consistent with flatness is viewed as supporting 
evidence for inflation. The expected deviations from 
flatness are typically at or below the level 
of the initial curvature perturbations at the end of 
inflation, $|\ok| \lesssim 10^{-5}$ (e.g., \cite{Kashlinsky94a,Bucher95a}). 
The ultimate precision with which the curvature may be determined 
from observations is limited by cosmic variance and 
model selection considerations to $\sigma(\ok)\sim 10^{-5}-10^{-4}$
\citep{Waterhouse08a,Vardanyan09a}.
While there are some theories of inflation in which the 
present value of the curvature is large enough to be potentially observable
without excessive fine tuning of the initial conditions of inflation
\citep{Gott82a,Ellis91a,Sasaki93a,Bucher95a,Linde95a,Hawking98a,Gratton02a,Uzan03a,Linde03a,Lasenby05a,Freivogel06a},
a detection of nonzero curvature would challenge 
at least the simplest inflationary theories.

The strongest observational bounds on curvature are presently 
at the percent level, i.e.\ $\sigma(\ok)\sim 0.01$.
The main limits on curvature come from measurements of 
angular diameter distances in the CMB at $z\sim 1000$ and 
BAO at $z < 1$.
However, these constraints rely on assuming a simple model 
for the dark energy such as a cosmological constant.
More precise and more model-independent measurements of the 
spatial curvature would provide valuable tests of theories 
of inflation.

%########################################
\subsection{Distances}
\label{sec:dist}

\begin{figure}[t]
\centerline{\includegraphics[width=3.5in]{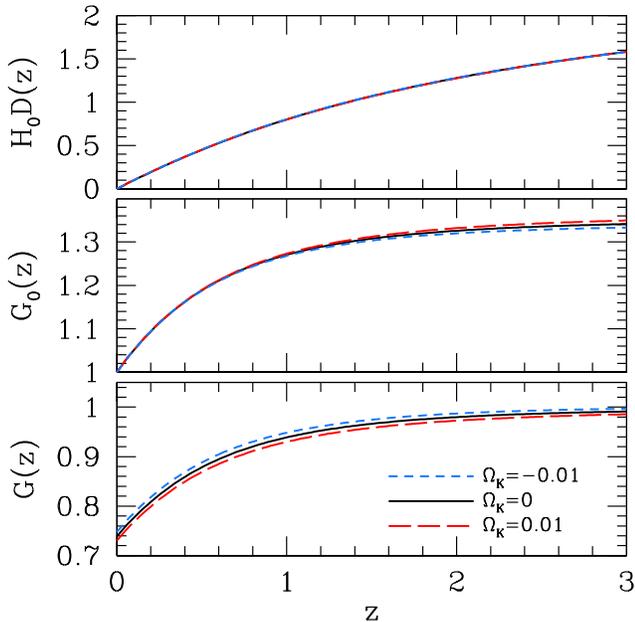}}
\caption
[Distance and growth evolution for flat, open, and closed \lcdm\ models.]
{
Comoving distance (\emph{top}), 
growth relative to $z=0$ (\emph{middle}), and growth 
relative to recombination (\emph{bottom}) vs.\ redshift 
for flat, open, and closed models. For all three models, 
$\om=0.24$ and $h=0.73$.
In the top panel, the 3 curves are indistinguishable in $H_0D(z)$.
}
%\vspace{.5cm}
\label{plot:dgfid}
\end{figure}

In a flat universe, the comoving distance $\tchi$ to an object at redshift $z$ 
obtained by integrating over the comoving radial coordinate in 
the FRW metric is
\begin{equation}
\tchi(z) = \int_0^z \frac{dz'}{H(z')}.
\label{eq:dist}
\end{equation}
More generally, for universes with nonzero spatial curvature 
the comoving distance is
\begin{equation}
D(z) = \frac{1}{\fcurv} S_{\rm K}\left[\fcurv 
\tchi(z) \right],
\label{eq:angdist}
\end{equation}
where $\fcurv \equiv (|\ok|H_0^2)^{1/2}$ is the inverse of 
the curvature radius of the universe, and 
$S_{\rm K}(x) = x$ for a flat universe, $\sinh x$ for an open 
universe, and $\sin x$ for a closed universe.
Distances therefore depend on both the expansion rate, $H(z)$, 
and geometry, $\ok$.
The distances at low redshifts for three models with varying 
spatial curvature are plotted in Fig.~\ref{plot:dgfid}.
Note that the curvature dependence is very weak at low redshifts, 
but high-redshift distances, e.g. the distance to recombination, 
are more sensitive to the geometry of the universe.

% ****************************************
% ****************************************

The main probes of distances we will consider here are Type Ia 
supernovae (SNe), standardizable candles whose average
magnitudes are related to distances as
\begin{equation}
m(z) = 5\log[H_0\dlum(z)]  + \scrm,
\label{eq:snmag}
\end{equation}
where $\scrm = M-5\log(H_0/{\rm Mpc}^{-1})+25$ 
combines the unknown absolute magnitude of
the supernovae $M$ and Hubble constant $H_0$, both of which only affect 
the overall normalization of the SN distance--redshift relation.
Since it involves unknown parameters that do not affect the 
measured evolution of distances with redshift, 
$\scrm$ is a ``nuisance'' parameter that is generally 
marginalized in a cosmological analysis of SN data.

Because the distance normalization is unknown, SN data determine 
relative distances, but not the absolute scale of the distance--redshift 
relation. However, SNe at low $z$ can constrain the normalization 
since $\lim_{z\to 0}H_0 D(z)=z$ under reasonable assumptions about 
the evolution of $H(z)$ in the recent past.\footnote{For example, 
assuming that there was not a sudden large transition in the dark 
energy equation of state at $z\lesssim 0.01$ (e.g., see \citetalias{Mortonson09a}.)}
Then for low-redshift SNe, the average magnitude from 
\eqn{snmag} is $m(z)\approx 5\log z+\scrm$ which provides an 
estimate of $\scrm$.

We will also consider the angular diameter distance constraint
from the acoustic scale 
of the CMB. The main effects of dark energy and curvature on the CMB 
enter through the distance to recombination at 
$z_*\approx 1089$ \citep{Spergel03a} and the matter density $\omhh$, which 
affect the angular scale and amplitude of the acoustic peaks \citep{Hu05a}. 
Dark energy and curvature also affect the large-scale CMB anisotropies 
through the integrated Sachs-Wolfe effect, but the information available 
is limited due to cosmic variance on those scales and the resulting 
constraints are relatively weak. We will therefore neglect this information 
in the curvature forecasts.

Angular diameter distances can also be determined by measuring baryon 
acoustic oscillations in the matter power spectrum 
in the plane transverse to the line of sight. BAO can be a powerful 
probe of absolute distances, but incomplete redshift coverage and 
the need for wide redshift bins make the technique less suitable than 
Type Ia SNe as a primary source of the distance information for predictions of 
the growth evolution. However, BAO data can provide complementary 
constraints on curvature through other means (e.g., \cite{Knox06a}).

%#######################################
\subsection{Growth of linear perturbations}
\label{sec:grow}

The growth of linear matter perturbations obeys 
\begin{equation}
\ddot{\delta}+2H\dot{\delta}-4\pi G_N\rho_{\rm m} \delta = 0
\label{eq:grow1}
\end{equation}
where $\delta \equiv \delta\rho_{\rm m}/\rho_{\rm m}$ and
overdots are derivatives with respect to time $t$.
We assume here and throughout this work that
general relativity is valid and that the dark energy is 
smooth on the relevant scales so that additional terms in 
the growth equation describing 
the clustering of dark energy can be neglected.

Equation~(\ref{eq:grow1}) can be written in terms of $G\propto (1+z)\delta$ as
\begin{eqnarray}
\frac{d^2G}{d\ln a} &+& \left(4+\frac{d\ln H}{d\ln a}\right)\frac{dG}{d\ln a}
\label{eq:growth}\\ 
&+& \left[3+\frac{d\ln H}{d\ln a}-\frac{3}{2}\om(z)\right]G = 0,\nonumber
\end{eqnarray}
where $\om(z)=\om H_0^2 (1+z)^3/H^2(z)$.
The growth variable $G$ is constant in a universe that 
contains only matter, so it is nearly constant at high redshifts 
during matter domination.
We normalize the growth functions to $\delta(z=0) = 1$ and $G(z\to \infty)=1$ 
so that
\begin{equation}
(1+z)\delta(z) = \frac{G(z)}{G(z=0)} \equiv G_0(z).
\label{eq:grownorm}
\end{equation}
Figure~\ref{plot:dgfid} shows $G_0(z)$ and $G(z)$ for three models with 
different values of the spatial curvature.

From Eqs.~(\ref{eq:angdist}) and~(\ref{eq:growth}) one can see that distances 
depend on both the expansion rate and geometry, while growth 
depends only on the expansion rate (and $\om H_0^2$ which is well determined 
by CMB data). Combinations of distance and growth information 
with similar redshift coverage therefore determine
the geometry of the universe with reduced dependence on the expansion rate.

Measurements of cluster abundances in a range of redshift bins 
can probe the growth evolution at low redshifts,
determining $G_0(z)$.
Clusters can also constrain $G(z=0)$ to probe high-redshift changes 
in the growth evolution due to massive neutrinos or early dark energy 
by comparing the growth determined by low-$z$ clusters 
with the predicted growth extrapolated to low redshifts from 
measurements of the CMB power spectrum amplitude at $z\sim 1000$ 
(e.g., see \cite{Rozo09a}).  
Ref.~\cite{Vikhlinin09a} contains current examples of both
types of measurements using an X-ray cluster sample.

Assuming that dark matter halos of mass $M$ host galaxy 
clusters with the same mass, 
the cluster abundance depends on cosmology primarily 
through the halo mass function $dn/dM$ and the comoving volume 
element in a solid element $d\Omega$ and redshift slice $dz$,
\begin{equation}
\frac{dV}{d\Omega~dz} = \frac{D^2(z)}{H(z)}.
\label{eq:volume}
\end{equation}
The mass function describing the comoving density of 
dark matter halos of mass $M$ at redshift $z$ can be written as
\begin{equation}
\frac{dn}{dM} = \frac{\om \rhoc}{M}~\frac{d\ln\sigma^{-1}(M,z)}{dM}~f(\sigma(M,z)).
\label{eq:massfn}
\end{equation}
Here $\sigma^2(M,z)$ is the variance of linear matter density 
perturbations,
\begin{equation}
\sigma^2(M,z) =  \left[\frac{G_0(z)}{1+z}\right]^2 
\int d\ln k ~\Delta^2(k) W^2(kR(M)),
\label{eq:var}
\end{equation}
where $R(M)=[3M/(4\pi\om\rhoc)]^{1/3}$, $\Delta^2(k)$ is the 
dimensionless power spectrum of linear matter perturbations at $z=0$ 
with comoving  wavenumber $k$,
and $W(kR)=3j_1(kR)/kR$ is the Fourier transform of a spherical top-hat 
window function with radius $R$.

The function $f(\sigma)$ in \eqn{massfn} parametrizes the mass 
function in a way that is relatively independent of 
redshift and cosmological parameters.
The dependence on $\sigma$ is exponential as $\sigma\to 0$, 
and since $\sigma(M,z)\propto G_0(z)$, 
the abundance of massive clusters is exponentially 
sensitive to the growth function.

Additional details about the cluster abundance
can be found in Sec.~\ref{sec:datadep2}, where we 
describe the dependence of curvature estimates from SN, CMB, 
and cluster data on the modeling of the cluster growth 
information.

%#####################################
\subsection{Expansion rate}

The Friedmann equation gives the expansion rate as determined by 
the evolution of the density of various components:
\begin{eqnarray}
H(z) &=& H_0 \Bigl[ \om (1+z)^3 + \orad (1+z)^4 \nonumber\\
&&\qquad+\frac{\rhode(z)}{\rhoc} + \ok (1+z)^2 \Bigr]^{1/2},
\label{eq:hz}
\end{eqnarray}
where $\rhode$ is the dark energy density and $\om$, $\orad$, and $\ok$ 
are the present matter density, radiation density, and effective 
curvature density, respectively, in units of 
the critical density. Here we 
generalize the dark energy phenomenology 
to allow arbitrary evolution of the dark energy density.
For a general time-dependent dark energy equation of state $w(z)$, the dark energy 
density evolves as
\begin{equation}
\rhode(z) = \rhoc \ode \exp\left[3 \int_0^z dz' \frac{1+w(z')}{1+z'}\right],
\end{equation}
where $\ode = 1-\om-\orad-\ok$ is the 
present fraction of dark energy.

The data we consider for curvature forecasts only have the ability to constrain 
the detailed dark energy evolution at low redshifts ($z<\zmax$ where 
$\zmax\sim 1.5$), 
so we only allow complete freedom in $\rhode(z)$ at late times.
In models with $w(z)\sim -1$, the dark energy fraction decreases rapidly 
with increasing redshift and therefore the exact form of the high-redshift 
dark energy evolution is unimportant; for example, for flat \lcdm, 
$\ode(z_*)\sim 10^{-9}$.
However, models have been proposed in which the dark energy
remains a significant fraction of the total density even at high $z$, 
such as scalar field models that track the density of the 
dominant matter and radiation components at early times 
\citep{Ratra88a,Ferreira97a,Steinhardt98a}.
To account for such possibilities, we parametrize 
early dark energy using an effective constant equation of 
state 
\begin{equation}
w(z>\zmax) = \winf,
\end{equation}
as in \citetalias{Mortonson09a}.
At $z>\zmax$, the dark energy density is
\begin{equation}
\rhode(z) = \rhode(\zmax)\left(\frac{1+z}{1+\zmax}\right)^{3(1+\winf)}.
\end{equation}
Although this may not be a realistic model for early dark energy, 
it should be sufficient to absorb at least small effects of 
early dark energy on curvature estimates.
Dark energy that behaves like a cosmological constant at $z>\zmax$ 
would have $\winf=-1$ and tracking models have $\winf=0$.
Other values of $\winf$ may provide an effective description of 
other types of models, e.g.\ tracking models that transition to 
$w<0$ at $z\gg \zmax$ can be approximated by a constant 
equation of state in the range $-1<\winf<0$.
We examine how well this approach works in the context of 
different early dark energy models (as well as models with massive 
neutrinos, which have effects that are indistinguishable 
below the neutrino free streaming scale from early dark energy
in the observables considered here) in Sec.~\ref{sec:cosmdep}.

Since $\winf$ only specifies the redshift evolution of the 
dark energy density at $z>\zmax$ but not its normalization, 
the density at some redshift is required to completely 
describe the early dark energy model.
It is convenient to use the fraction of dark energy at 
$\zmax$, $\ode(\zmax)$, for this purpose.
One way to estimate this quantity from SN data is to 
differentiate $\tchi(z)$ at $\zmax$, since
from Eqs.~(\ref{eq:dist}) and~(\ref{eq:hz}) we can obtain
\begin{equation}
\frac{\rhode(\zmax)}{\rhoc} = \emax^2-\sum_{i\ne {\rm DE}}\Omega_i
(1+\zmax)^{3(1+w_i)},
\label{eq:odemax}
\end{equation}
where $\emax = H(\zmax)/H_0 = H_0^{-1}(\partial z/\partial \tchi)|_{z=\zmax}$.

\vspace{1cm}
\subsection{Data for forecasts} \label{sec:data}

In this section, we summarize the main data assumptions we use for 
the curvature forecasts presented in Sec.~\ref{sec:curv}. These 
data include future Type Ia supernova and CMB 
observations as probes of 
distances, and X-ray cluster abundances as a probe of the linear 
growth history. The characteristics of the 
former set of distance data match those assumed in
\citetalias{Mortonson09a}.

The supernova sample is taken to match the planned redshift 
distribution for the SuperNova/Acceleration Probe (SNAP) \citep{SNAP}, 
which covers redshifts $0.1<z<1.7$. In addition, we assume 
a sample of 300 low-redshift SNe at $0.03<z<0.1$.
The intrinsic SN magnitude dispersion is taken to be $\sigma_{\rm stat}=0.15$, 
and the systematic error is modeled as 
$\sigma_{\rm sys} = 0.02[(1+z)/2.7]$ in redshift bins of width $\Delta z=0.1$.
Then the uncertainty in relative distances from SNe in a $\Delta z=0.1$ bin 
with $N$ SNe is 
\begin{equation}
\sigma_{\ln H_0 D}=0.2\ln 10 \sqrt{N^{-1} \sigma_{\rm stat}^2 +
\sigma_{\rm sys}^2}.
\end{equation}

The CMB distance priors we use are modeled on the specifications 
for the recently-launched Planck satellite \citep{Planck}. 
As in \citetalias{Mortonson09a}, we describe the CMB data 
with a 2D Fisher matrix $F^{\rm CMB}$
including the distance to recombination, $D(z_*)$, and 
the physical matter density, $\omhh$. The elements of the 
covariance matrix $C^{\rm CMB} = (F^{\rm CMB})^{-1}$ are 
$C^{\rm CMB}_{xx} = (0.0018)^2$, $C^{\rm CMB}_{yy} = (0.0011)^2$, 
and $C^{\rm CMB}_{xy} = -(0.0014)^2$, where 
$x = \ln [D(z_*)/{\rm Mpc}]$ and $y = \omhh$.

In addition to the distance constraints from SN and CMB data, 
in some cases we will consider the effect of additional priors.
For the MCMC analysis, we use the priors of \citetalias{Mortonson09a}
which correspond to constraints from currently available data, 
including an $11\%$ $H_0$ prior 
from HST Key Project data \citep{Freedman01a}, a $3.7\%$ BAO measurement 
of $D(z=0.35)$ from SDSS \citep{Eisenstein05a}, 
and a $2.5\%$ upper limit on the fraction of early 
dark energy at recombination from the WMAP temperature 
angular power spectrum \citep{Doran07a}.
For the growth reconstruction forecasts, we also consider future 
priors including a $1\%$ measurement of $H_0$ \citep{Greenhill09a}
and a $1\%$ upper limit on the dark energy fraction at 
recombination from CMB data, $\ode(z_*)$.
A stronger limit of $\sim 0.2\%$ on the early dark energy fraction 
may be obtainable with future observations of CMB lensing in the context of 
specific parametrizations of early dark energy 
\citep{dePutter09a,Hollenstein09a}, but here we adopt a more 
conservative prior to allow for the possibility that the limit
may weaken upon including more general early dark energy behavior.

Each of these constraints supplementing the SNAP and Planck data 
is implemented as a Gaussian prior with mean 
equal to the value in the fiducial cosmology.
We will see in Sec.~\ref{sec:curv} that for a variety of 
cosmological models, particularly those in which the deviations in 
the high-redshift density evolution from the concordance model 
are mild, these additional priors are unnecessary for curvature estimates 
from future distance and growth data sets.

For growth forecasts, we consider constraints on the amplitude 
and redshift evolution of galaxy cluster abundances using  
observations from the proposed International X-ray Observatory (IXO) 
\citep{Vikhlinin09b}. The IXO is projected to obtain $1-2\%$ 
measurements of the growth function $G_0(z)$ in $\Delta z=0.1$ 
bins over $0<z<2$, assuming that distances and the expansion rate 
are effectively fixed by other data sets such as SNe. 
When combined with the expected $1\%$ 
measurement by Planck of the amplitude of scalar fluctuations from the 
CMB power spectra, the X-ray cluster data should also provide 
a $\sim 1-2\%$ measurement of $G(z)$ at $z<2$. 

We approximate the cluster growth information 
from IXO (and Planck for comparison of the cluster abundance 
and CMB power spectrum amplitudes) by a Gaussian likelihood 
with uniform uncertainties in $G_0(z)$ at $z\leq \zmax$ in 
$\Delta z=0.1$ bins and $G(\zmax)$, using $\zmax=1.5$ 
corresponding to the maximum redshift at which distances 
are well constrained by SNe.
Our forecasts in Sec.~\ref{sec:curv} assume optimistic 
$1\%$ growth uncertainties 
from IXO as the default assumption, but in Sec.~\ref{sec:datadep2}
we also examine how 
curvature constraints weaken for more pessimistic assumptions and
consider how degeneracies 
between distances and growth in the cluster observables may 
affect the predicted curvature constraints.

\vspace{.4cm}

%%%%%%%%%%%%%%%%%%%%%%%%%%%%%%%%%%%%%%%%%%%
\section{The geometric degeneracy}
\label{sec:degen}

In this section, we present examples of the degeneracy between 
curvature and dark energy evolution in the distance--redshift relation, 
and show how growth information can break this degeneracy.

The geometric degeneracy 
can take a variety of forms, depending on the dark energy modeling and the 
available data. For example, even for a cosmological constant $\Lambda$ there 
is a degeneracy between $\olam$ and $\ok$ if the only input is the 
distance to recombination from the CMB (e.g., \cite{Bond97a,Zaldarriaga97a,Efstathiou99a}).
Adding more data breaks the degeneracy for the cosmological constant model, but 
the degeneracy persists for more complex dark energy models. 
Taken to an extreme, even if one has exact measurements of the distance--redshift 
relation over the entire history of the universe, there is still a 
degeneracy if we allow arbitrary evolution of the dark energy density 
with redshift. This general form of the geometric degeneracy is what we 
focus on here.

The degeneracy in distance data 
is apparent if we differentiate \eqn{angdist} and use \eqn{hz} to 
solve for the dark energy density:
\begin{equation}
\frac{\rhode(z)}{\rhoc}=\frac{1+\ok [H_0^{\rm fid}D^{\rm fid}(z)]_{\rm SN}^2}{(\partial [H_0^{\rm fid}D^{\rm fid}(z)]_{\rm SN}
/\partial z)^2} - \sum_{i\ne {\rm DE}}\Omega_i
(1+z)^{3(1+w_i)} ,
\label{eq:geomdegen}
\end{equation}
assuming that SN data provide relative distance measures $[H_0^{\rm fid}D^{\rm fid}(z)]_{\rm SN}$ 
for some fiducial cosmology.
[Note that this generalizes \eqn{odemax}.]
Thus for any value of the spatial curvature, there exists some 
dark energy evolution that matches a given set of distance measurements.

The ability of dark energy to match distances for any value of $\ok$
is weakened if we introduce some mild restrictions on the dark 
energy evolution, for example requiring that the dark energy density 
be nonnegative. Observational constraints 
beyond the distance--redshift relation can also limit the 
possible dark energy behavior; for example, a dark energy fraction of 
a few percent or more at recombination would distort the CMB temperature power spectrum 
in ways that are not observed in WMAP data \citep{Doran07a,Zahn03a,Wright07a}.

In practice, however, the redshift coverage of data is limited, 
so degenerate cosmological models only need to match distances at 
the redshifts where we can actually measure distances.
Within the 
redshift ``gaps'' in our observations, the dark energy evolution can deviate from 
\eqn{geomdegen} and still satisfy all available observational constraints.

\begin{figure}[t]
\centerline{\includegraphics[width=3.7in]{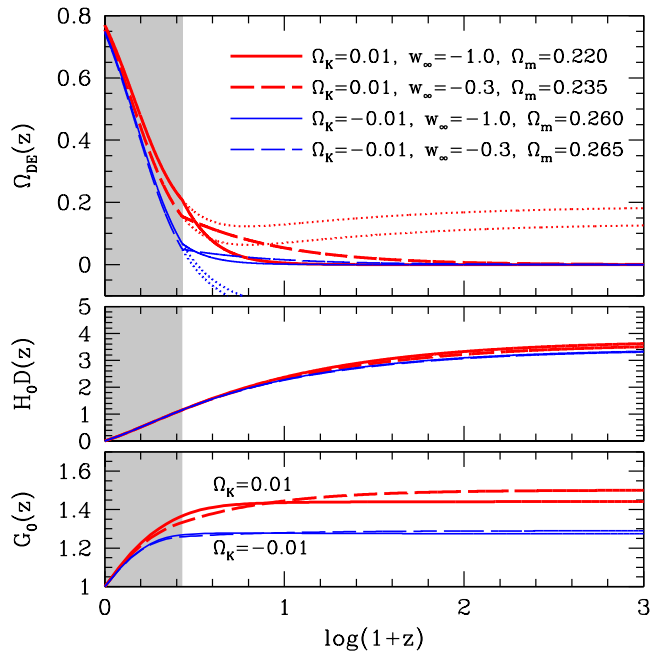}}
\caption
[Examples illustrating the geometric degeneracy between 
curvature and dark energy dynamics.]
{Geometric distance degeneracy between curvature and dark energy 
dynamics. Curves show the dark energy fraction (\emph{top panel}), 
relative distances (\emph{middle panel}), and 
growth relative to $z=0$ (\emph{bottom panel}) for models with  
distance--redshift relations that are degenerate with a fiducial flat 
\lcdm\ model with $\om=0.24$ and $h=0.73$. Thick red curves 
correspond to open models with $\ok=0.01$, and thin blue curves 
are closed models with $\ok=-0.01$. 
The solid and dashed curves only match the relative 
distances at $z<1.7$ (\emph{shaded region}) 
and the CMB distance at $z\sim 1000$, but not distances at intermediate redshifts 
where the dark energy equation of state is assumed to be 
constant, $w(1.7<z<z_*)=\winf$.
Dotted curves in the top panel show the dark energy evolution 
that would be required to match relative distances of the 
fiducial model at \emph{all} redshifts.
Parameters of each model are given in the top panel legend;
for each model, $h$ is set so that $\omhh$ is the same as 
in the fiducial model.
}
%\vspace{.8cm}
\label{plot:geomdegen}
\end{figure}

Figure~\ref{plot:geomdegen} shows the dark energy fraction, 
distance, and growth of four example models chosen to illustrate 
these points. 
These models all have nonzero curvature (two open and 
two closed), but $\om$, $h$, and $\winf$ are adjusted so that they all 
match relative SN distances 
$H_0D(z)$ at $z<\zmax$ 
with $\zmax=1.7$ (see middle panel 
of Fig.~\ref{plot:geomdegen}) and CMB constraints on 
$D(z_*)$ and $\omhh$ for a flat \lcdm\ model with $\om=0.24$ 
and $h=0.73$.

% ****************************************
% ****************************************

For each model, a dotted line shows how the 
dark energy evolution would extend to $z>\zmax$ if we required the 
models to match $H_0D(z)$ at all distances in the range $0<z<z_*$ 
(but not the CMB absolute distance); for the closed models 
this would require a negative dark energy density, $\ode(z)<0$, at high $z$,
and for the open models $\ode(z_*)\sim 10-20\%$ which violates 
current CMB constraints. However, due to the lack of observational 
constraints at $\zmax<z<z_*$, not only can we find models with 
nonzero curvature that match the SN and CMB data of the fiducial 
flat \lcdm\ model, but there is in fact a set of degenerate models 
for each value of $\ok$ with different combinations of the matter 
density and early dark energy parameters.

Specifically, to match an observed distance to recombination
$[D^{\rm fid}(z_*)]_{\rm CMB}$ as well as SN distances,
the constraint that these models must satisfy [in 
addition to \eqn{geomdegen} at $z<\zmax$] is
\begin{eqnarray}
\fcurv \int_{\zmax}^{z_*} \frac{dz}{H(z)} &=& 
S_K^{-1}\{\fcurv [D^{\rm fid}(\zmax)]_{\rm CMB}\}\nonumber\\
&-&S_K^{-1}\{\fcurv H_0^{-1} [H_0^{\rm fid}D^{\rm fid}(z_*)]_{\rm SN}\}.
\label{eq:geomdegen2}
\end{eqnarray}

The bottom panel of Fig.~\ref{plot:geomdegen} shows that despite being 
degenerate in the SN and CMB distance data, these models are distinct 
in their growth evolution for different values of the spatial curvature.
This separation based on curvature is mostly independent of the 
$\{\om,\winf\}$ values.
This indicates that growth observations can break the degeneracy 
between $\ok$, $\om$, and $\winf$ that is present in SN and CMB data 
and provide model-independent information about the curvature.

Note that a precise independent measurement of $H_0$ (or, equivalently, 
combinations of SN relative distances with absolute distances, e.g.\ from 
BAO) determines the value of $\om$ since the CMB precisely constrains
$\omhh$. This removes one parameter from the degeneracy described 
above, so the remaining degeneracy is between curvature and 
dark energy only with the matter density fixed. The freedom to match 
observations for any value of $\ok$ is thereby 
reduced, leading to constraints on curvature even in the absence of 
growth information. We discuss these constraints in relation to 
the curvature estimates from distances and growth in Sec.~\ref{sec:curv}.

%\vspace{.5cm}

%%%%%%%%%%%%%%%%%%%%%%%%%%%%%%%%%%%%%%%%%%%
\section{MCMC method}
\label{sec:mcmc}

For both the MCMC analysis presented in this section and 
the growth reconstruction method (Sec.~\ref{sec:analytic}), 
it is convenient to view the constraints on curvature from distances and 
growth in the following way. First, we assume that 
the distance--redshift relation is precisely measured at 
low $z$ by SNe and at high $z$ by the CMB. Given this 
distance data and assuming the validity of GR, 
we can use either the MCMC likelihood analysis or the 
analytic growth reconstruction technique to compute a 
\emph{predicted} growth history that is consistent with 
the measured distances.
The main sources of uncertainty in 
the relation between measured distances and predicted 
growth are curvature and early dark energy (or massive neutrinos), 
with curvature primarily affecting the growth relative to 
$z=0$ [$G_0(z)$] and early dark energy affecting the 
growth relative to high $z$ [$G(z)$]
(\citetalias{Mortonson09a}). 
Because of this dependence, \emph{measurements} of $G_0(z)$ 
yield constraints on curvature when they are compared with the predicted growth history,
and measurements of $G(z)$ 
help constrain deviations from matter domination at high $z$.

Computing the distances and growth functions 
for each MCMC sample of the cosmological parameter space
is straightforward as it only 
requires using Eqs.~(\ref{eq:dist}) and~(\ref{eq:angdist}) to 
obtain the distance--redshift relation and solving \eqn{growth} 
for the growth. We will see in Sec.~\ref{sec:analytic} that 
the growth reconstruction scheme is somewhat more complicated 
to implement.

One of the main advantages of using the MCMC approach is 
that the estimation of parameter uncertainties is also straightforward: 
as long as the MCMC samples have converged to a stationary 
distribution approximating the joint posterior probability 
of the parameters (Sec.~\ref{sec:predict}), the marginalized probability for 
curvature can be obtained by binning the samples based on the 
value of $\ok$. However, the parameter chains sometimes 
converge quite slowly, especially for very general 
parametrizations of dark energy combined with curvature.
External priors like those described in Sec.~\ref{sec:data} 
can improve MCMC convergence in some cases.

A drawback of the MCMC analysis is that although we 
are trying to obtain model-independent results, we must still 
specify some model for the dark energy evolution. 
To provide a general parametrization at low $z$, 
we use several principal components of the dark energy equation of state 
as described in the following section.
However, there is still some degree of unavoidable dependence on 
dark energy priors. In the absence of strong growth constraints, 
the choice of priors can influence the curvature constraints 
as we will see in Sec.~\ref{sec:curv}. The growth reconstruction 
method of Sec.~\ref{sec:analytic} does not require specifying a 
parametrization for the dark energy equation of state and therefore 
suffers less from such effects.

%##############################################
\subsection{Growth predictions from distances}
\label{sec:predict}

The MCMC method of predicting 
the linear growth of perturbations is based on computing the growth functions 
of a large sample of cosmological models that fit observed 
distances well. The range of growth functions spanned by 
these models constitutes a prediction for the growth evolution 
based on distances. This procedure is described in detail by
\citetalias{Mortonson09a}, and here we present a summary.

% ****************************************
\begin{figure}[b]
\centerline{\includegraphics[width=3in]{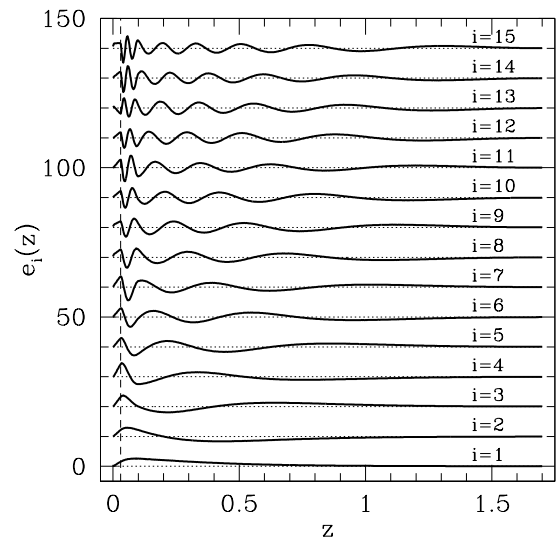}}
\caption
[Principal components of the dark energy equation of state for 
SNAP SN and Planck CMB forecasts.]
{The first 15 principal components of $w(z)$ 
for SNAP and Planck (increasing variance from
  bottom to top), with 500 redshift bins between $z=0$ and
  $\zmax=1.7$ and $\wfid=-1$. The vertical dashed line shows the minimum redshift of the data
  assumed for computing the PCs, $\zminsn = 0.03$.
  The PCs are offset vertically
  from each other for clarity with dotted lines showing the zero point for each
  component.
}
%\vspace{.3cm}
\label{plot:pcs}
\end{figure}
% ****************************************

Two main ingredients are required: a parametrization of 
cosmological models and a description of how well these models 
fit the observed distances. Since we are concerned about 
possible degeneracies between dark energy and curvature, we 
want to allow a wide variety of dark energy behavior.
This is accomplished using a basis of 
principal components (PCs) for the dark energy equation of state $w(z)$ 
to parametrize general dark energy evolution below a redshift $\zmax$:
\begin{equation}
w(z) - \wfid(z) = \sum_{i} \alpha_i e_i(z),
\label{eq:pcstow}
\end{equation}
where the PCs are ordered according to how well they are measured 
by a particular combination of data.
The PCs are eigenvectors of the Fisher matrix for the distance 
data, taken here and in \citetalias{Mortonson09a} 
to be the SNAP SN sample, CMB distance data from Planck, and 
current priors on $H_0$, $D(z=0.35)$, and $\ode(z_*)$ 
as described in Sec.~\ref{sec:data}.
Figure~\ref{plot:pcs} shows the redshift dependence of the 
15 lowest-variance PCs.

The Fisher matrix for the PCs is computed at some fiducial dark energy 
model specified by the equation of state $\wfid(z)$, usually 
taken to be a cosmological constant.
The redshift range of the PCs is the same as the range 
of the SN data, with maximum redshift $\zmax=1.7$.
The PCs are normalized as
\begin{equation}
\sum_{i=1}^{\nzpc} [e_i(z_j)]^2 = \sum_{j=1}^{\nzpc} [e_i(z_j)]^2 = \nzpc,
\label{eq:norm}
\end{equation}
where $\nzpc$ is the number of redshift bins,
so that the shapes of the PCs are roughly independent of the chosen 
bin width.

The highest-variance principal components have a negligible effect 
on observable distances and growth due to their rapid oscillation in 
redshift. We therefore truncate the sum in Eq.~(\ref{eq:pcstow}) at 
15 PCs, found by \citetalias{Mortonson09a} 
to be a sufficient number for a complete representation of
the effects of dark energy variation at $z<\zmax$ on the 
distance and growth observables.
The dark energy description is completed by specifying the 
high-redshift evolution through the constant effective 
equation of state $w(z>\zmax)=\winf$.
Besides varying $\winf$ and the $w(z)$ PC amplitudes in the MCMC
analysis, we also include  
$\om$, $H_0$ (parametrized through the combination $\omhh$), 
and $\ok$ as MCMC parameters, so there are 19 parameters in all.

Using top-hat priors on the PC amplitudes, one can restrict the value of $w(z)$ 
to a particular range, conservatively erring on the side of allowing too many 
models rather than too few.  
The priors corresponding to the range $w_{\rm min}<w<w_{\rm max}$
are $\alpha_i^{(-)} \leq \alpha_i \leq \alpha_i^{(+)}$,
where
\begin{eqnarray}
\alpha_i^{(\pm)} &\equiv& \frac{1}{2\nzpc} \sum_{j=1}^{\nzpc}
[(\wmin+\wmax-2\wfid)e_i(z_j) \nonumber\\
&& \pm (\wmax-\wmin) |e_i(z_j)|\,],\label{eq:prior1}
\end{eqnarray}
assuming constant $\wfid(z)$ (\citetalias{Mortonson09a}).\footnote{Similar top-hat priors are derived in Ref.~\cite{Mortonson08a} 
in the context of principal components of the reionization history.}
For example, requiring $-1\leq w\leq 1$ 
for quintessence models places strong limits on the allowed 
PC amplitudes. In the interest of keeping the dark energy evolution 
as general as possible, here we will only consider the weakest 
priors on $w$ used in \citetalias{Mortonson09a} corresponding 
to $-5\leq w\leq 3$.

By varying the cosmological parameters, computing the likelihood of 
each model for the assumed distance data sets, and 
using the Metropolis-Hastings criterion for deciding whether or not 
to accept a proposed step in the parameter space, the resulting 
set of parameter combinations trace out the joint posterior 
probability of the parameters (e.g., \cite{Christensen01a,Kosowsky02a,Dunkley05a}).
To determine when the number of MCMC samples is large enough that 
the parameter chains have converged to the posterior distribution, 
we run 4 independent chains and require that the Gelman-Rubin statistic 
satisfy $R-1\lesssim 0.01$, indicating that the variance of the mean 
value of a parameter between different chains is much smaller than 
the variance within a single chain \citep{Gelman92a}.

This MCMC procedure produces a variety of cosmological models that 
fit the fiducial distance data reasonably well. 
For each of these models, we compute the growth history using 
\eqn{growth}.
The distribution of the resulting growth functions then forms the 
prediction for growth from distances.

% ****************************************
\begin{figure}[t]
\centerline{\includegraphics[width=3.4in]{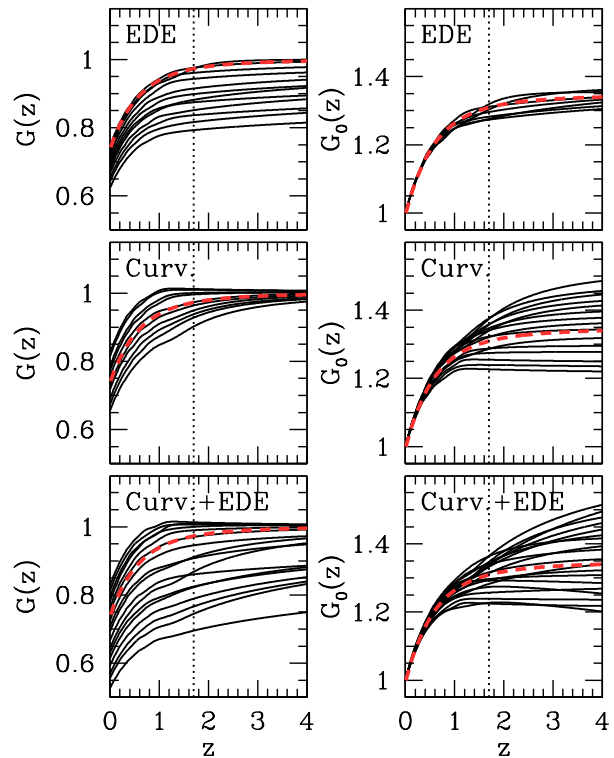}}
\caption
[Growth functions of MCMC models matched to SN and CMB distances, 
allowing freedom in early dark energy, curvature, or both.]
{Growth functions of MCMC samples 
with general dark energy equation of state variations at 
$z<1.7$ within the range $-5\leq w \leq 3$, including
either early dark energy (EDE) at $z>1.7$ ($\winf \ne -1$; \emph{top}), 
curvature ($\ok\ne 0$; \emph{middle}), or both (\emph{bottom}). 
The left panels show growth relative to 
early times, and the right panels show growth relative to the present.
Dashed red curves show growth in the fiducial flat \lcdm\ model 
($\om=0.24$, $h=0.73$).
Samples are selected randomly from those with likelihoods 
satisfying $\Delta \chi^2 \leq 4$, but for visual clarity we plot 
samples that are approximately evenly spaced in $G(z=0)$ (\emph{left}) 
or $G_0(z=4)$ (\emph{right}).
The dotted vertical line in each panel marks the division between 
the low-$z$ and high-$z$ dark energy descriptions at $z=1.7$.
}
%\vspace{.5cm}
\label{plot:gzmax}
\end{figure}
% ****************************************

Plotting sets of predicted growth functions from 
MCMC analyses with different degrees 
of freedom in the cosmological models reveals 
how distance-matched growth functions depend on curvature and 
early dark energy. Figure~\ref{plot:gzmax} shows the growth evolution of 
selected MCMC samples in chains with early dark energy varying and  
curvature fixed to $\ok=0$, 
curvature varying and early dark energy fixed to $\winf=-1$, 
or variation in both curvature and early dark energy
(in addition to variation in the low-$z$ 
dark energy equation of state via principal components). 
Early dark energy mainly affects the growth 
amplitude relative to high redshift, $G(z)$, with very little effect 
on the shape of the growth evolution at low $z$ characterized by 
$G_0(z)$. On the other hand, curvature strongly influences $G_0(z)$ 
but has less of an effect on $G(z)$: for a fixed distance--redshift 
relation, open (closed) models have larger (smaller) $G_0(z)$ than 
a flat universe. This is the same effect that we see in the 
geometric degeneracy examples in Fig.~\ref{plot:geomdegen}.

This difference between the effects of curvature and 
early dark energy supports the claim
that measurements of $G_0(z)$, combined with distance measurements,
can constrain
curvature with little dependence on early dark energy or other 
high-redshift phenomena. Measurements of $G(z)$ can be used to 
place limits on any residual effects that early dark energy might 
have on these curvature estimates.

One interesting feature of the spread of the distance-matched $G_0(z)$ evolution 
due to curvature is that it tells us
how precise growth measurements must be to improve on 
model-independent curvature constraints. In particular, for 
the forecasts shown here, the uncertainty in the predicted 
growth at $z\sim 1$ (at $68\%$~CL) is about 
$5-10\%$ (\citetalias{Mortonson09a}),\footnote{Note that the uncertainty in 
growth predictions depends on the priors assumed in addition to the 
SN and CMB data forecasts, in this case taken from current $H_0$, BAO, and CMB 
data as described in Sec.~\ref{sec:data}.} 
so any direct measurements of the growth evolution that are less 
precise than this will not appreciably reduce the 
uncertainty in $\ok$.
For example, current cluster measurements 
of growth are not yet precise enough to significantly 
reduce the uncertainty in curvature from current distance constraints
\citep{Vikhlinin09a}.

As observed in \citetalias{Mortonson09a}, MCMC estimates of $\ok$ from 
distance data are dependent on 
$w(z)$ priors if ``phantom'' dark energy models ($w<-1$) are allowed.
This dependence is essentially a volume effect related to 
the large volume of degenerate $w<-1$ models that 
are correlated with $\ok<0$ \citep{Upadhye05a}.
Precise growth measurements help reduce this dependence on 
dark energy priors as we will see in Sec.~\ref{sec:curv}.

%#####################################################
\subsection{MCMC estimates of curvature}
\label{sec:mcmcest}

The procedure described in the previous section produces 
chains of parameter combinations that match the 
SN and CMB distance data, as well as the additional priors.
To obtain forecasts for $\ok$ from distance and growth data, 
we need to add the growth information from 
measured cluster abundances.

The first step for including the growth constraints is to 
simply compute the growth evolution for each MCMC sample. 
With a growth history associated with each distance-matched 
MCMC sample, we then use importance sampling of the parameter 
chains \citep{Lewis02a}
to reweight the samples according to the 
growth likelihood described in Sec.~\ref{sec:data}.
The posterior probability for $\ok$ 
from the distance and growth forecasts is then computed as usual by 
marginalizing over the other MCMC parameters.
In Sec.~\ref{sec:curv}, we will describe the resulting curvature 
constraints and compare them with the forecasts from the 
growth reconstruction method.

%\vspace{.3cm}

%%%%%%%%%%%%%%%%%%%%%%%%%%%%%%%%%%%%%%%%%%%
\section{Growth reconstruction}
\label{sec:analytic}

The MCMC approach in the previous section used a general 
parametrization of the dark energy equation of state to 
predict growth evolution from measured distances by 
searching for $w(z)$ satisfying the distance constraints and computing 
the corresponding growth evolution. 
However, it is possible to go from distances to growth 
directly without using an intermediary like $w(z)$.
Here we will first summarize this growth reconstruction method 
and then show how it can be used to study 
model-independent curvature constraints. 
Additional details about the procedure we use for growth 
reconstruction from simulated SN and CMB data are provided 
in Appendix~\ref{sec:mcsims}.

The growth reconstruction equations derived by 
Alam, Sahni, and Starobinsky \cite{Sahni06a,Alam08a}
express the amplitude of linear perturbations $\delta$ as a 
function of the comoving distance assuming spatial flatness, 
$\tchi$ [\eqn{dist}].
For convenience, we define a dimensionless comoving distance, $\chi \equiv H_0\tchi$.\footnote{Note that in Ref.~\cite{Alam08a} this quantity is called $E$, but here we use 
the notation $\chi$ instead to avoid confusion with the 
common definition $E(z) = H(z)/H_0$.}

Starting from Eq.~(\ref{eq:grow1}) for the linear growth function, 
we can rewrite the equation in terms of $\chi$ by
using $d/dt=-H_0(1+z)d/d\chi$ and writing the matter density as 
$\rho_{\rm m} = 3\om H_0^2/(8\pi G_N) (1+z)^3$
 to get
\begin{equation}
H_0^2 (1+z)[(1+z)\delta']' - 2H_0H(1+z)\delta'-\frac{3}{2}H_0^2\om(1+z)^3\delta = 0,
\end{equation}
where primes denote derivatives with respect to $\chi$.
Since $d\chi/dz = H_0/H$, we can replace $H$ by $H_0~dz/d\chi = H_0(1+z)'$.
After dividing by $H_0^2(1+z)^3$, this yields
\begin{equation}
\frac{[(1+z)\delta']'-2(1+z)'\delta'}{(1+z)^2} = \frac{3}{2}\om\delta.
\end{equation}
The left hand side of this equation is equal to 
$[(1+z)\delta''-(1+z)'\delta']/(1+z)^2$, which can be written as
\begin{equation}
\left(\frac{\delta'}{1+z}\right)' = \frac{3}{2}\om \delta.
\label{eq:growrec0}
\end{equation}

Integrating Eq.~(\ref{eq:growrec0}) over $\chi$ leads to the main 
growth reconstruction equation \citep{Sahni06a,Alam08a}, which is an integral equation
for $\delta(\chi)$:
\begin{eqnarray}
\delta(\chi) &=& 1+\delta_0^{\prime} \int_0^\chi d\chi_1 [1+z(\chi_1)] \label{eq:growrec1}
\\
&& + \frac{3}{2}\om\int_0^\chi d\chi_1 [1+z(\chi_1)] \int_0^{\chi_1} d\chi_2
~\delta(\chi_2),\nonumber
\end{eqnarray}
where $\delta_0'$ is the derivative of $\delta(\chi)$ at $z=0$, 
and the growth function is normalized to $\delta(z=0)=1$.
Solving for the growth function involves making an initial guess for 
$\delta(\chi)$ and plugging it into the right hand side of 
Eq.~(\ref{eq:growrec1}), taking the resulting $\delta(\chi)$ from 
the left hand side and plugging it back in to the right hand side, 
and repeating until the solution has converged.

As noted in Ref.~\cite{Alam08a}, 
\eqn{growrec1} only requires integration of observed quantities, so
the growth reconstruction method is more stable to the presence of scatter in the data 
than other methods that use derivatives of the observed distance--redshift 
relation.
Differentiation uses information from the data over only a small range of 
redshift or $\chi$, whereas an integral from $0$ to $\chi$ uses data over 
the entire range which helps average out statistical noise. For $\chi$ near 
$0$ the number of data points used is still small, but the 
reduced statistical power can be offset by having smaller intrinsic 
uncertainties at lower redshifts.

By differentiating \eqn{growrec1} and evaluating it at $\chi = \chi(\zmax)$, 
we can shift the boundary condition on the derivative of $\delta(\chi)$ 
from $z=0$ to $z=\zmax$:
\begin{equation}
\delta'_0 = \frac{\dpmax}{1+\zmax}-\frac{3}{2}\om \int_0^{\chi(\zmax)}d\chi~\delta(\chi), \label{eq:shift}
\end{equation}
where $\dpmax=\delta'(\chi(\zmax))$.
Then \eqn{growrec1} can be rewritten as
\begin{eqnarray}
\delta(\chi) &=& 1+\frac{\dpmax}{1+\zmax} \int_0^\chi d\chi_1 [1+z(\chi_1)] \label{eq:growrec2}
\\
&& - \frac{3}{2}\om\int_0^\chi d\chi_1 [1+z(\chi_1)] \int_{\chi_1}^{\chi(\zmax)} d\chi_2
~\delta(\chi_2).\nonumber
\end{eqnarray}
This is the form of the growth reconstruction equation that we will 
use for the curvature forecasts.

One advantage of setting the boundary condition for $\delta'(\chi)$ 
at $\chi(\zmax)$ instead of $\chi=0$ is that
$\dpmax$ depends mainly 
on the assumed cosmology at high redshifts, whereas $\delta'_0$ depends on 
not only the high-$z$ assumptions but also the low-$z$ SN data constraints and 
iterative growth solution.
Also, setting $\dpmax$ makes it easier to 
ensure that $\delta(\chi)$ is smooth at $\chi(\zmax)$ by requiring the 
same derivative there for both the fiducial model at $z>\zmax$ and 
the reconstructed growth function at $z<\zmax$.
Rather than setting the value of the $\delta'(\chi)$ boundary 
condition using an integral over $\delta(\chi)$ as suggested in 
Ref.~\cite{Sahni06a}, here we use an approximate analytic form for the 
high-redshift growth function valid when matter is the dominant 
component.

Solving \eqn{growrec2} for $\delta(\chi)$ requires specifying the 
function $z(\chi)$ and three parameters: $\ok$, $\om$, and $\dpmax$.
Some of these inputs to the growth reconstruction are set by the SN and CMB 
data constraints, while others are free parameters.
Because of this link between the distance data sets and the growth 
reconstruction, some of the scatter in the SN and CMB observations 
will propagate to uncertainties in the reconstructed growth. 
We compute the mean values of reconstructed growth observables 
and their uncertainties using Monte Carlo simulations of the SN and 
CMB data, modeled on SNAP and Planck as for the MCMC method.

The following steps summarize the procedure for these Monte Carlo 
simulations. Additional details about each step are provided in 
Appendix~\ref{sec:mcsims}.

\begin{enumerate}
\item Assume values for the curvature and any additional parameters 
describing the high-redshift cosmology; here we use $\winf$ for 
parametrizing early dark energy. 
The output of the Monte Carlo simulations will be the conditional probability for the growth observables given $\ok$ and $\winf$.
\item Draw a realization of the SN data [$H_0D(z)$] for the fiducial 
cosmology and estimate $z(\chi)$ from the data. Given an assumed value of $\ok$, we can invert \eqn{angdist} to get
\begin{equation}
\chi(z) = \frac{1}{\sqrt{|\ok|}} S_K^{-1}\left[
\sqrt{|\ok|} H_0 D(z) \right],
\end{equation}
where $S_K^{-1}(x)$ is $\sinh^{-1} x$ for an open universe,
$\sin^{-1} x$ for a closed universe, and $x$ if $\ok=0$.
To reduce bias in the estimated $\chi(z)$ relation, we take the 
maximum redshift for the growth reconstruction to be $\zmax=1.5$, 
slightly lower than the maximum SN redshift of $z=1.7$ that 
was used as $\zmax$ in the MCMC method of Sec.~\ref{sec:mcmc} (see 
Appendix~\ref{sec:mcsims} for details).
\item Draw a realization of the CMB data [$D(z_*)$ and $\omhh$] and use 
\eqn{geomdegen2} along with the assumed values of $\ok$ and $\winf$ 
to compute the value of $\om$ required by the CMB parameters.
\item Steps $1-3$ fix the cosmological model at $z>\zmax$, so we 
can compute $G(\zmax)$ for this cosmology as one of the growth observables, and 
use the approximate high-$z$ growth solution of Appendix~\ref{sec:growapprox}
[\eqn{dpzmax}] to set $\dpmax$ for the growth reconstruction.
\item Solve \eqn{growrec2} to find the reconstructed growth, $\delta(\chi)$, 
for the particular realization of SN and CMB data. Use the $\chi(z)$ relation 
from step 2 to express this solution as $\delta(z)$ 
or $G_0(z)$.
\end{enumerate}

Repeating these steps for many realizations of the distance data 
produces a distribution of the growth observables ${\bf g}$, 
which include $G(\zmax)$ from step 4 and $G_0(z)$ from step 5, for the chosen 
values of $\ok$ and $\winf$. This procedure can be carried out 
at several different values of $\ok$ and $\winf$ to map out 
the conditional probability $P({\bf g}|\ok,\winf)$.
This probability describes the growth predictions from distance data 
in the context of the growth reconstruction method.

% ****************************************
\begin{figure}[t]
\centerline{\includegraphics[width=3.4in]{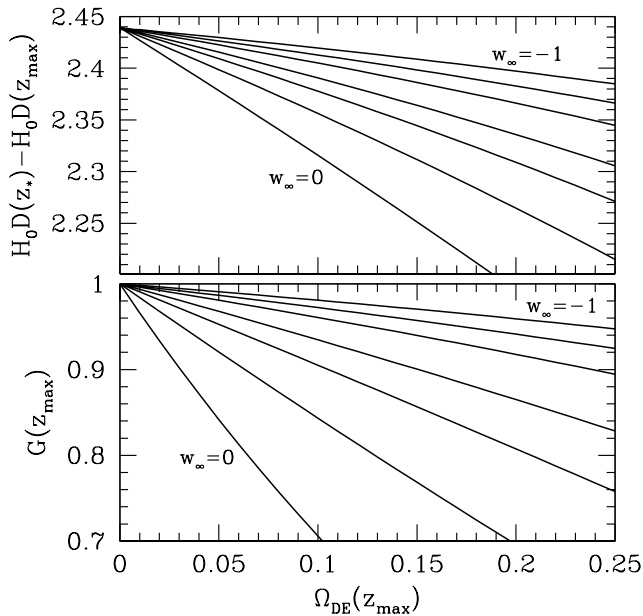}}
\caption
[Effect of early dark energy on distance and growth.]
{Effect of early dark energy on distance and growth.
\emph{Top panel}: Difference between the distance to recombination 
and the distance to $\zmax=1.5$. \emph{Bottom panel}: Growth at 
$\zmax$ relative to recombination. Early dark energy models are 
parametrized by the fraction of dark energy at $\zmax$ and 
$\winf=w(z>\zmax)$. From top to bottom in each panel, 
$\winf = -1$, $-0.7$, $-0.5$, $-0.3$, $-0.2$, $-0.1$, and $0$.
Flat \lcdm\ is assumed here with $\om=0.24$ and $h=0.73$.
}
%\vspace{1cm}
\label{plot:gzmaxede}
\end{figure}
% ****************************************

Note that in step 3, there will generally be degeneracies between $\ok$, 
$\om$, and early dark energy in the CMB constraints, increasing 
the uncertainty in the model-independent estimate of curvature.
Fortunately, measurements of growth relative to high $z$ 
can reduce this uncertainty.
The top panel of Figure~\ref{plot:gzmaxede} shows the effect of various early 
dark energy models on the difference between the distance to recombination 
and the distance to $\zmax$. 
Both $\winf>-1$ and $\ode(\zmax)\gtrsim 0.05$ are required for 
the distances to be significantly affected.
As the bottom panel of Fig.~\ref{plot:gzmaxede}
shows, the growth relative to high redshift,
$G(\zmax)$, is sensitive to early dark energy in a way that is 
similar to how the distances depend on early dark energy. 
Therefore, precise measurement of the 
growth at $\zmax$ relative to growth at early times (for example, 
by comparing the normalization of growth from cluster abundances with 
the amplitude of CMB power spectra) can constrain the effect of 
early dark energy on high-$z$ distances, thereby reducing the 
uncertainty in the $\om$--$\ok$ relation from the CMB constraints. 
Although $G(z)$ predicted from distances 
is relatively insensitive to curvature in a direct sense (see Fig.~\ref{plot:gzmax}), 
it provides an important complementary constraint to the 
curvature estimates from distances and $G_0(z)$ that reduces 
model dependence.

% ****************************************
\begin{figure}[tb]
\centerline{\includegraphics[width=3.2in]{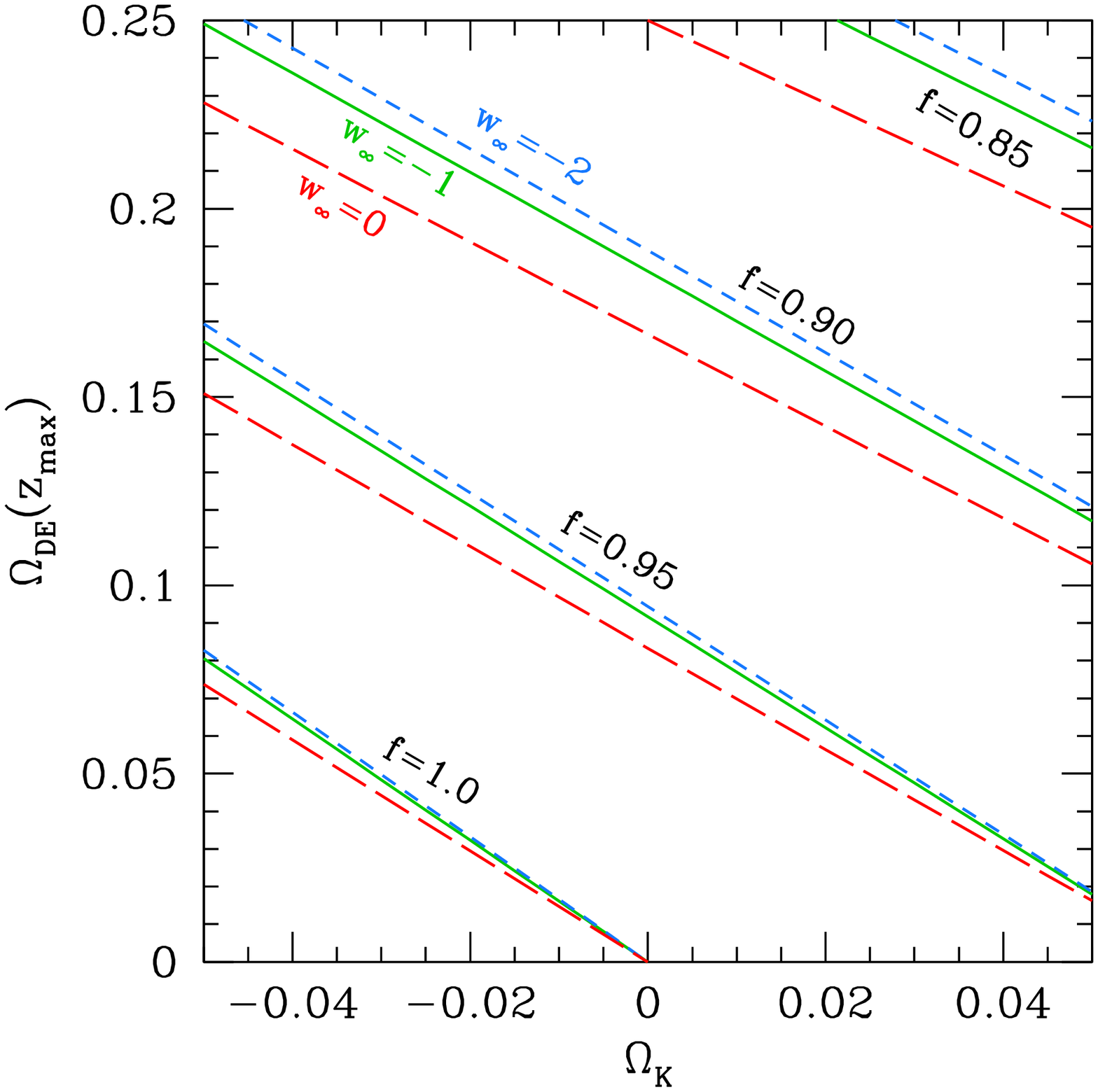}}
\centerline{\includegraphics[width=3.2in]{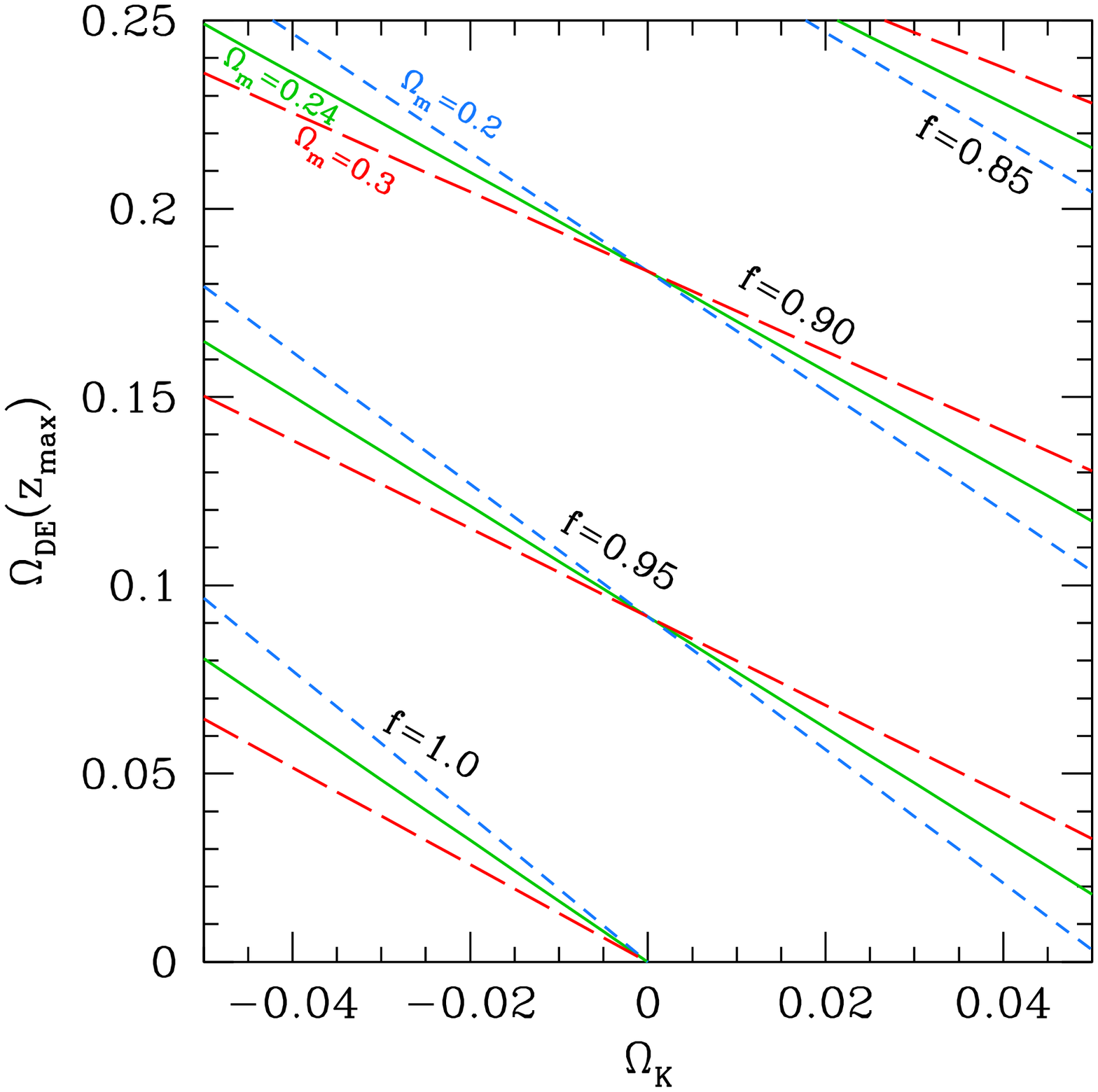}}
\caption
[Dependence of the growth rate at $z=1.5$ on curvature and
early dark energy.]
{Contours of the growth rate $f$ at $\zmax=1.5$ as a function of 
$\ok$ and $\ode(\zmax)$, for three choices of $\winf$ at fixed
$\om=0.24$ (\emph{top}) and three choices of $\om$ at fixed $\winf=-1$ (\emph{bottom}).
}
%\vspace{.5cm}
\label{plot:fcontours}
\end{figure}
% ****************************************

The approximation for $\dpmax$ used in step 4, which is described in 
Appendix~\ref{sec:growapprox}, assumes that the curvature and 
early dark energy fractions at high $z$ are small enough that 
the growth can be written as a perturbation to the matter-dominated 
solution where $G(z)$ is constant.
The dependence of $\dpmax$ on early dark energy turns out to be 
fairly weak. Figure~\ref{plot:fcontours} shows the value of the 
growth rate $f = d\ln \delta/d\ln a = 1+d\ln G/d\ln a$ 
at $\zmax$ as a function of 
curvature and the fraction of dark energy at $\zmax$. 
Changing $\winf$ only slightly shifts the contours of $f$, 
especially for models that satisfy CMB constraints on the 
dark energy fraction at recombination.
Therefore, at fixed curvature $\dpmax/\delta(\zmax)=-(1+\zmax)^{-1}E(\zmax)f(\zmax)$ mainly depends on 
$\ode(\zmax)$, which can be estimated from the SN data using \eqn{odemax}.
Likewise, for curvature at the level of a few percent or less, 
uncertainty in $\om$ does not 
strongly affect $\dpmax$.

In summary, the growth reconstruction method as outlined above 
requires choosing values for two parameters: the curvature $\ok$ 
and the early dark energy equation of state $\winf$.
All other quantities are estimated from the distance information
provided by measurements of SNe and the CMB.
Given this distance data, the end result of the growth 
reconstruction procedure is a prediction for the growth evolution 
for each $\{\ok,\winf\}$ pair, $P_d({\bf g}|\ok,\winf)$.

\begin{figure}[t]
\centerline{\includegraphics[width=3.5in]{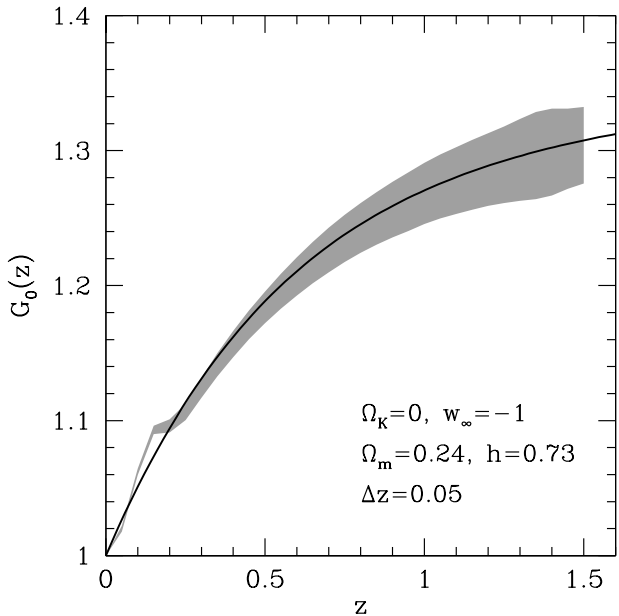}}
\caption
[Example of the growth function reconstructed from 
distance measurements for a flat \lcdm\ model.]
{Growth reconstruction for $\ok=0$ and $\winf=-1$ where 
the fiducial cosmology is flat \lcdm\ with $\om=0.24$ and $h=0.73$.
The shaded band is the 68\% CL region for $G_0(z)$ reconstructed 
from simulated SNAP and Planck data, and the true growth evolution is 
plotted as a solid curve.
The redshift bin width for the reconstruction is $\Delta z=0.05$.
}
%\vspace{.8cm}
\label{plot:growrecex}
\end{figure}

Figure~\ref{plot:growrecex} shows an example of the reconstructed 
growth evolution from the distance data for the fiducial flat \lcdm\ 
model, assuming $\ok=0$ and $\winf=-1$.
In the lowest redshift bins, the growth reconstruction is biased due to the 
relatively small number of SNe in the SNAP distribution at low $z$, 
but over most of the redshift range the reconstruction is in 
good agreement with the true growth function.

% ****************************************
% ****************************************

%#####################################################
\subsection{Growth reconstruction estimates of curvature}
\label{sec:growrecest}

By combining the predicted growth from distances $P_d({\bf g}|\ok)$
with measurements of the growth observables ${\bf g}$, 
we can obtain an estimate of $\ok$ from the growth reconstruction 
method:\footnote{For notational compactness, 
in this section we suppress dependence on additional 
parameters varied in the growth reconstruction analysis 
such as $\winf$, but in general, these should 
appear along with $\ok$ wherever there is dependence on curvature.}
\begin{equation}
P(\ok) = \int d{\bf g}~P_d(\ok,{\bf g})P_g({\bf g}),
\label{eq:pok1}
\end{equation}
where $P_d(\ok,{\bf g}) = P_d({\bf g}|\ok)P_{\rm prior}(\ok)$
is the joint probability for curvature and growth observables
from the growth reconstruction from distance data 
as described in the previous section combined with any additional 
prior information about $\ok$, and 
$P_g({\bf g})$ represents the constraints on the growth 
observables from probes of the growth history such as clusters.
We use subscripts $d$ and $g$ to indicate constraints coming from 
distance and growth data, respectively.

Assuming that the Monte Carlo simulations produce 
conditional probabilities for growth observables at
fixed $\ok$ that can be approximated as a multivariate Gaussian, we can 
write
\begin{equation}
P_d({\bf g}|\ok) = \frac{1}{(2\pi)^{n/2}(\det {\bf F}_d)^{-1/2}}
\exp\left( -\frac{1}{2}\delta {\bf g}^{\rm T}{\bf F}_d \delta {\bf g}\right),
\end{equation}
where ${\bf F}_d$ is the Fisher matrix for the $n$ growth observables 
from the simulated distance data, 
computed by inverting 
the covariance matrix ${\bf C}_d$ from the growth reconstruction 
Monte Carlo simulations, 
and $\delta {\bf g}\equiv {\bf g}-\bar{\bf g}_d$ is the deviation of the 
growth observables from the mean growth reconstruction solution.
Note that both ${\bf F}_d$ and $\bar{\bf g}_d$ depend on the 
value of $\ok$ (and additional parameters such as $\winf$).

If the growth observations are also well approximated by a 
multivariate Gaussian with covariance matrix ${\bf C}_g$,  
Fisher matrix ${\bf F}_g = {\bf C}_g^{-1}$, and 
mean values $\bar{\bf g}_g$ (assumed to be equal to the true 
growth history of the fiducial model), then the 
posterior probability for $\ok$ in \eqn{pok1} is
\begin{eqnarray}
P(\ok)&\propto& P_{\rm prior}(\ok)~[\det({\bf I}+{\bf F}_d^{-1}{\bf F}_g)]^{-1/2} \label{eq:pok2}\\
&\times& \exp\left[\frac{1}{2}{\bm \Delta}^{\rm T}{\bf F}_g({\bf F}_d+{\bf F}_g)^{-1}{\bf F}_g{\bm \Delta} -
\frac{1}{2}{\bm \Delta}^{\rm T}{\bf F}_g{\bm \Delta}\right],\nonumber
\end{eqnarray}
where ${\bm \Delta}\equiv \bar{\bf g}_d-\bar{\bf g}_g$ is the difference 
between the average growth evolution predicted from distances at 
a particular assumed value of $\ok$ and the true growth evolution, and
${\bf I}$ is the $n\times n$ identity matrix.
We describe the resulting forecasts for curvature in the next section.

\vspace{.3cm}

%%%%%%%%%%%%%%%%%%%%%%%%%%%%%%%%%%%%%%%%%%%%%%%%%%%%%%%%%%%%%%%%%%%%%
\section{Model-independent curvature constraints} \label{sec:curv}

In this section, we use the techniques for combining distance 
and growth measurements described in Sections~\ref{sec:mcmc} 
and~\ref{sec:analytic} to obtain forecasts for spatial curvature 
constraints from the simulated SN, CMB, and cluster data of Sec.~\ref{sec:data}.

The accuracy of curvature estimates depends not only on 
the assumed characteristics of the distance and growth data sets, but 
also on the fiducial, ``true'' cosmological model assumed for the 
forecasts. We begin in Sec.~\ref{sec:flcdm} 
with the simple case in which 
the fiducial model is flat \lcdm, and use this example to compare 
forecasts from the MCMC and growth reconstruction methods. 
In Sec.~\ref{sec:cosmdep}, we generalize to other fiducial cosmologies 
that are more or less consistent with constraints from current data.
Tests of the dependence of these results on modeling of 
the SN, CMB, and cluster data sets are described in 
Sections~\ref{sec:datadep1} and~\ref{sec:datadep2}.
Finally, Sec.~\ref{sec:compare} compares these distance and growth constraints 
on curvature with other model-independent tests of curvature.

%#####################################################
\subsection{Flat \lcdm}
\label{sec:flcdm}

\begin{figure}[t]
\centerline{\includegraphics[width=3.5in]{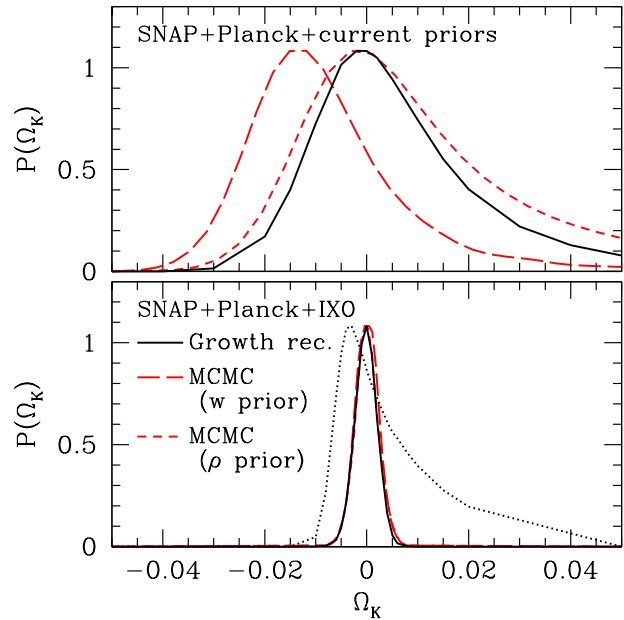}}
\caption
[Comparison of MCMC and growth reconstruction 
curvature forecasts from distances and growth 
for the flat \lcdm\ fiducial cosmology.]
{Comparison of MCMC and growth reconstruction 
curvature forecasts for the flat \lcdm\ fiducial cosmology.
\emph{Top panel:} combining \emph{current} 
 priors on $H_0$, $D(z=0.35)$, and $\ode(z_*)$
with SN and CMB forecasts based on SNAP and Planck, respectively.
\emph{Bottom panel:} combining SNAP and Planck forecasts with 
cluster forecasts based on IXO as a probe of the growth evolution.
For comparison, the dotted curve shows a forecast for SNAP, Planck, 
and future $1\%$ measurements of $H_0$ and $\ode(z_*)$ (without 
growth constraints).
}
%\vspace{.4cm}
\label{plot:compare}
\end{figure}

We begin by comparing MCMC and growth reconstruction constraints 
on $\ok$ in the context of a flat \lcdm\ fiducial cosmology, 
taking the parameters to be $\om=0.24$ and $h=0.73$.
Figure~\ref{plot:compare} shows curvature forecasts for two different 
sets of data. Both cases include distance constraints from SNe modeled
after SNAP and CMB data based on Planck. On their own, these data 
sets would place almost no limits on curvature, assuming that 
general forms of the dark energy evolution are allowed.

% ****************************************
% ****************************************

In the top panel of Fig.~\ref{plot:compare}, the SN and CMB data are 
supplemented by weak priors based on current data: an $11\%$ $H_0$ prior 
from HST Key Project data \citep{Freedman01a}, a $3.7\%$ BAO measurement 
of $D(z=0.35)$ from SDSS \citep{Eisenstein05a}, 
and a $2.5\%$ upper limit on the fraction of early 
dark energy at recombination from the WMAP temperature spectrum \citep{Doran07a}.
We adopt these priors here for consistency with the MCMC analysis 
of model-independent growth predictions in \citetalias{Mortonson09a}.

Using only the current priors in addition to the SN and CMB data, 
curvature is determined with an accuracy of $\sigma(\ok)\sim 0.02$.
However, the results depend on the analysis method. In particular, 
the MCMC constraints on $\ok$ depend strongly on the assumed 
priors on the dark energy parameters (\citetalias{Mortonson09a}).
For priors that are flat in the $w(z)$ principal component 
amplitudes [$P(\alpha_i)=$ constant], 
the curvature constraint is biased toward $\ok<0$.
Taking alternate priors which are instead flat in the 
\emph{density} associated with each principal component at 
$\zmax=1.7$, 
\begin{equation}
\frac{\rho_i(\zmax)}{\rho_i(0)} = \exp\left[ 3\alpha_i
\int_0^{\zmax} dz \frac{e_i(z)}{1+z}\right],
\end{equation}
the resulting curvature constraint is less biased, but has a long tail 
at $\ok>0$.
The curvature estimate from the growth reconstruction method 
using the same data is
more consistent with the flat-density prior constraint from the 
MCMC analysis, but has a slightly narrower distribution.

The lower panel of Fig.~\ref{plot:compare} shows the resulting curvature 
forecasts when we drop the current priors on $H_0$, BAO distance, 
and the early dark energy fraction, and instead combine the 
SN and CMB distances with $1\%$ growth function measurements from
IXO clusters.
Using growth to break the geometric degeneracy in distances, the 
curvature constraint improves to an accuracy of 
$\sigma(\ok)=0.0022$. Not only is this a much stronger constraint 
on curvature than without the growth information, but it is also 
significantly less dependent on the parameter estimation methodology.
The MCMC constraints using different types of dark energy priors 
are consistent with each other and unbiased, and the 
forecast from growth reconstruction agrees with the MCMC results.

Since we are taking a forecast for future data as the source 
of growth information, it is not quite fair to compare this future distance plus 
growth constraint with the curvature estimate from \emph{current}
Hubble constant, BAO, and early dark energy priors.  
We use those priors only to illustrate the possible dependence of 
estimates that do not use growth information on dark energy priors or other details 
of the analysis, and for comparison with the earlier MCMC 
results from \citetalias{Mortonson09a}. If we instead combine 
SNAP and Planck forecasts with \emph{future} 
priors on $H_0$ and $\ode(z_*)$, both with $1\%$ accuracy (see Sec.~\ref{sec:data}), 
the curvature estimate improves but remains weaker and more skewed than 
the distance plus growth forecasts (see bottom panel of Fig.~\ref{plot:compare}). 
We discuss the impact of future BAO measurements and other types of data 
on model-independent curvature constraints in comparison to  
constraints from distances and growth in Sec.~\ref{sec:compare}.

% ****************************************
\begin{figure}[t]
\centerline{\includegraphics[width=3.5in]{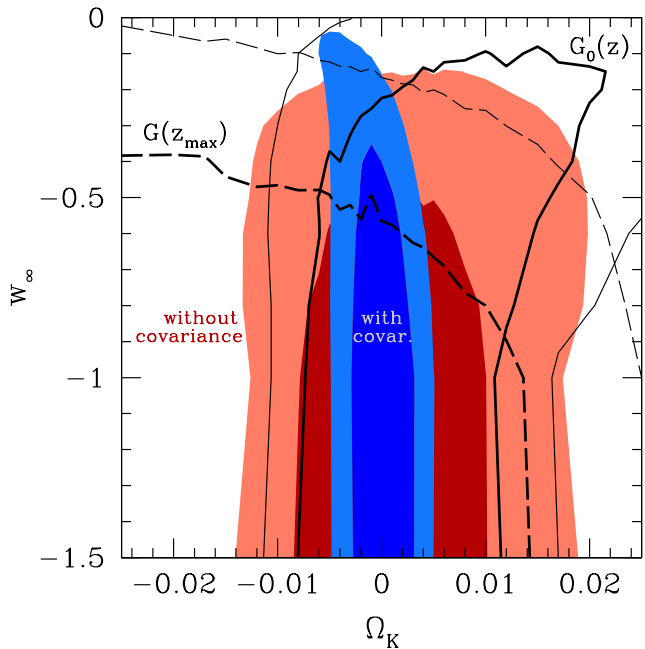}}
\caption
[Curvature and early dark energy forecast for 
the flat \lcdm\ fiducial model.]
{Forecasts in the $\ok$--$\winf$ plane for 
the flat \lcdm\ fiducial model, including various combinations 
of growth information in addition to SN and CMB distance constraints: 
$G(\zmax)$ (\emph{dashed contours}), 
$G_0(z)$ at $z\leq 1.5$ (\emph{solid contours}), or 
both (\emph{shaded blue contours}). 
The combined growth constraint \emph{without} the 
covariance between the predicted $G(\zmax)$ and $G_0(z)$
from the growth reconstruction is shown as shaded red contours.
Contours are plotted at $68\%$ CL (\emph{thick curves/dark shading}) and 
$95\%$ CL (\emph{thin curves/light shading}).
}
%\vspace{.2cm}
\label{plot:okwinf}
\end{figure}
% ****************************************

The MCMC and growth reconstruction forecasts show that
combinations of distances and growth have the potential to 
provide reliable, model-independent measurements of curvature 
with $\sim 0.2\%$ accuracy, at least in the context of a 
flat \lcdm\ cosmology. Before exploring the forecasts for 
this method for other fiducial cosmological models, we examine 
how different parts of the distance and growth data contribute 
to the curvature estimate.

Figure~\ref{plot:okwinf} shows the impact of different components of 
the assumed growth information from IXO cluster data on the
joint constraint on curvature and early dark energy.
Recall from Sec.~\ref{sec:data} that there are two types of 
growth information used here: $1\%$ measurements of the 
growth evolution at $0<z<\zmax$ that probe $G_0(z)$, and 
a $1\%$ measurement of the growth at $\zmax$ relative to the 
growth at recombination, $G(\zmax)$. The unshaded contours in 
Fig.~\ref{plot:okwinf} show the results of using each of these 
growth constraints separately (in addition to the SN and CMB data).

If the growth information consists only of $G_0(z)$ (i.e.\ just the 
relative evolution of the cluster mass function at low $z$ with 
unknown normalization relative to the CMB), the constraints on $\ok$ 
are a few times weaker than the combined growth constraints for 
$\winf\lesssim -1$. As $\winf$ 
approaches 0, the $G_0$ constraint shifts toward open models. 
As a result, $P(\ok)$ marginalized over $\winf$ is weakened even further 
and is skewed toward $\ok>0$.

On the other hand, growth information from $G(\zmax)$ only 
(e.g., comparing $\sigma_8$ from clusters and the CMB)
places very weak constraints on curvature, only significantly limiting 
the range of allowed open models. However, the information provided 
by $G(\zmax)$ is complementary to that from $G_0(z)$ since it cuts off 
the $G_0$ degeneracy at $\ok>0$ and $\winf\gtrsim -0.5$. By removing 
this degeneracy, the combined growth constraints are much less 
sensitive to the early dark energy parameters, resulting in a stronger, 
unbiased curvature estimate.

Moreover, the covariance between $G(\zmax)$ and $G_0(z)$ predictions 
from measured distances reduces the curvature uncertainty beyond 
what would be expected from simply combining the separate
$G(\zmax)$ and $G_0(z)$ constraints. As shown in Fig.~\ref{plot:okwinf}, 
the distance plus growth constraints on curvature with the $G(\zmax)-G_0(z)$
covariance removed by hand (red shaded contours) are a few times 
weaker than the full constraint including this covariance (blue 
shaded contours).

The covariance between the growth observables comes primarily from 
the CMB distance constraint. Matching the distance to recombination 
requires a balance between the low-$z$ and high-$z$ dark energy 
evolution. For example, increasing the dark energy density at low 
redshifts tends to decrease $D(z_*)$, and decreasing the high-redshift 
dark energy density can compensate for this shift. The CMB distance 
priors therefore anticorrelate the dark energy evolution at early and late 
times. This results in a positive correlation between the 
predicted values of $G_0(z)$ and $G(\zmax)$ from SN and CMB data. 
In the example above, the larger dark energy density at low $z$ 
enhances the suppression of the late-time growth of perturbations, which 
corresponds to higher values of $G_0(z)$. A smaller dark energy density 
at high $z$ results in less growth suppression at early times and 
therefore a higher value of $G(\zmax)$.

This positive correlation between $G_0$ and $G$ for growth reconstructed 
from distances is the opposite of the effect of curvature. Relative to a 
flat universe, an open geometry tends to suppress growth at both early 
and late times while a closed geometry has the opposite effect. 
This means that increasing $\ok$ increases $G_0(z)$ while decreasing 
$G(\zmax)$, and vice versa. The positive covariance between these two 
types of growth observables required by CMB constraints therefore 
leads to stronger limits on curvature.

To summarize, in the case of flat \lcdm\ the primary information 
about curvature from growth 
combined with distance data comes from $G_0(z)$, as expected based on 
the discussions in the previous sections. 
However, normalization of low-redshift 
growth relative to the high redshift of recombination can 
significantly improve the accuracy of the curvature estimate by 
reducing early dark energy uncertainties and through the covariance 
with the low-$z$ growth evolution required to match SN and CMB distances.

%#####################################################
\subsection{Dependence on cosmology}
\label{sec:cosmdep}

We now turn to curvature forecasts from distances and growth for 
fiducial cosmologies other than the flat \lcdm\ 
example of the previous section. Since the analytic growth 
reconstruction method is more efficient at exploring different 
assumptions about the true cosmology and properties of the data, 
we will primarily rely on that method for the forecasts in this section.

As the first test of dependence 
on the fiducial cosmology underlying the data, 
we consider flat \lcdm\ models with different parameters from 
those in the previous section. For a model with $\om=0.3$ and 
$h=0.7$, the constraint on curvature is almost identical to 
that for $\om=0.24$ and $h=0.73$. Thus variation of flat \lcdm\ 
parameters in the range allowed by current data does not significantly 
affect the curvature forecasts.

\begin{figure}[t]
\centerline{\includegraphics[width=3.5in]{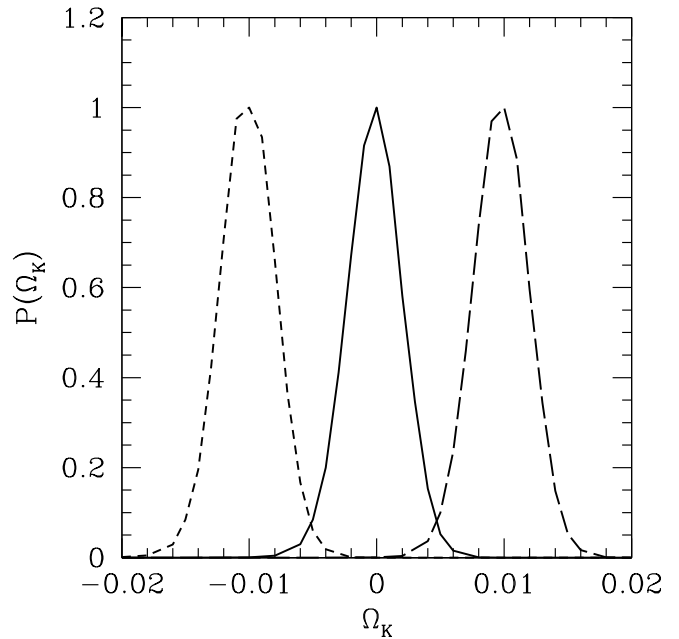}}
\caption
[Curvature forecasts for models 
with nonzero curvature.]
{Growth reconstruction forecast 
of the marginalized probability of $\ok$ for models 
with $\ok=0$ (\emph{solid}), $\ok=-0.01$ 
(\emph{short dashed}), and $\ok=0.01$ (\emph{long dashed}).
Each model is a \lcdm\ cosmology with $\om=0.24$ and $h=0.73$.
}
\vspace{.5cm}
\label{plot:pok1}
\end{figure}

Next, we consider models with different values of the curvature.
We saw in the previous section that if the universe is actually flat, 
future distance and growth probes can exclude at 95\% CL alternate models with 
$|\ok|\gtrsim 0.005$. It is also interesting to ask whether true nonzero
curvature could be detected (i.e.\ distinguished from $\ok=0$)
using this method.
Figure~\ref{plot:pok1} compares the forecast for the flat \lcdm\ 
model of the previous section with forecasts for an open model 
and a closed model, both with $|\ok|=0.01$ and 
the other parameters unchanged. The resulting constraints on curvature for 
the open and closed models are nearly identical to that for the flat 
model, except for being shifted to be centered on the true 
curvature. Therefore, distances and growth from SNAP, Planck, and 
IXO data enable percent-level curvature to be cleanly 
distinguished from flatness.

% ****************************************
% ****************************************

Curvature forecasts are similarly precise and unbiased for 
models with different dark energy evolution from \lcdm\ at 
low redshifts. For example, if we adopt the commonly-used 
parametrization of the dark energy equation of state
$w(z) = w_0 + w_a z/(1+z)$ \citep{Chevallier01a,Linder03a}
with $w_0=-0.8$ and $w_a=-0.5$, 
and keep all other parameters the same as in the fiducial flat 
\lcdm\ model, then the growth reconstruction analysis again 
returns an unbiased estimate of $\ok$ with $\sigma(\ok)\approx 0.002$.
Note that in this model, dark energy is negligible at high 
redshifts since $\lim_{z\to\infty}w(z)=w_0+w_a$ which is less than $-1$.

Although changing parameter values in the context of flat \lcdm\
and similar cosmological models
has little effect on the accuracy of the estimated curvature 
from future distances and growth, changing the fiducial model for 
the high-redshift universe can have more interesting 
consequences. 
We consider first early dark energy scenarios and then 
models with massive neutrinos.
In both cases, the $G(\zmax)$ constraint takes on a much larger 
role and constraints from $G_0(z)$ alone become unreliable.

% ****************************************
\begin{figure}[t]
\centerline{\includegraphics[width=3.5in]{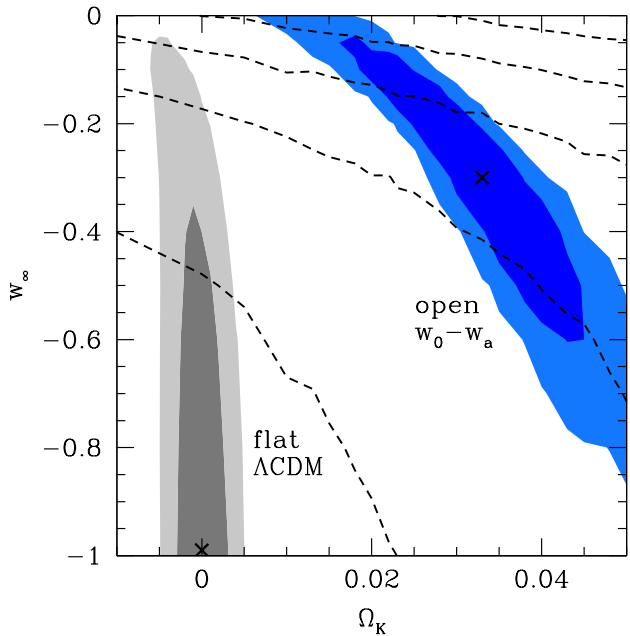}}
\caption
[Curvature forecast for an open universe with early dark energy.]
{Forecast in the $\ok$--$\winf$ plane for an open 
universe with early dark energy: $\ok=0.025$, $w_0=-1.08$, $w_a=1.02$ 
(\emph{shaded blue contours,right}).
The mean value of $G(\zmax)$ for the growth function reconstructed 
from the fiducial distances of this model is plotted with dashed, 
unshaded contours at $G(\zmax)=0.5, 0.6, 0.7, 0.8,$ and $0.9$, from 
top to bottom.
The flat \lcdm\ forecast from Fig.~\ref{plot:okwinf2} 
is plotted for comparison (\emph{shaded gray contours,left}). 
Crosses mark the best fit point for each model.
Shaded contours are plotted at $68\%$ and $95\%$ CL.
}
%\vspace{.3cm}
\label{plot:okwinf2}
\end{figure}
% ****************************************

Given the fact that current data are consistent with flat \lcdm,
models with significant early dark energy are more likely to 
be open than flat or closed since the geometric effect of 
negative spatial curvature can compensate for the reduced distance 
to recombination due to early dark energy (e.g., \cite{Ichikawa07a}).
Figure~\ref{plot:okwinf2} shows the SNAP+Planck+IXO constraints 
on a $w_0-w_a$ model with $w_0+w_a\approx 0$ to act as early dark energy 
and $\ok=0.025$ compared with the flat \lcdm\ model of the previous 
section.

For the open early dark energy model, the best fit to distance and growth 
constraints is near the true parameter values (the 
effective value of $w$ at $z>\zmax$ that matches the fiducial 
CMB distance and growth function normalization
is $\winf\approx -0.25$).
However, a significant 
degeneracy between curvature and early dark energy remains even with 
the combination of distance and growth data.
One consequence of this is that 
the curvature constraints marginalized over $\winf$ would be misleading;
for example, the shape of the $w_0-w_a$ model constraints 
in the $\ok-\winf$ plane
cause $P(\ok)$ to be biased high, 
although maximizing the likelihood over $\winf$ 
instead of marginalizing would reduce the apparent bias somewhat.
Furthermore, the marginalized $\ok$ 
constraints are weaker than the $0.2\%$ flat \lcdm\ estimate, 
despite the fact that the contours in the $\ok-\winf$ plane 
are of comparable width. 

Despite the loss of precision in the curvature estimates in the 
presence of early dark energy, the constraints from distances and 
growth for such models remain extremely interesting since they 
provide clear evidence for either large amounts of early dark energy 
or nonzero curvature, if not both. And, in fact, for models like 
this example with $\ok\gtrsim 0.02$ it is still possible to 
confidently exclude a flat universe independent of the assumptions
about early dark energy.

Unlike the models without early dark energy where the main curvature 
constraints come from distances plus $G_0(z)$, and $G(\zmax)$ has a 
lesser role (see Fig.~\ref{plot:okwinf}), for early dark energy 
models the $G(\zmax)$ constraint combined with distance data is 
the main source of both the early dark energy \emph{and} curvature 
constraints. In fact, curvature estimates using distances and 
measurements of $G_0(z)$ alone would indicate a preference for flatness and 
no early dark energy even if the true model was the $w_0-w_a$ example 
with $\ok=0.025$. Other types of curvature constraints, for example 
using a $1\%$ $H_0$ prior in addition to SNAP and 
Planck data, similarly fail to detect the nonzero curvature and early 
dark energy for such models. 
While a measurement of $G(\zmax)$ is a helpful additional 
constraint on models in which dark energy only becomes 
important at late times,  it 
is crucial for obtaining accurate constraints on models with 
early dark energy. Figure~\ref{plot:okwinf2} shows contours of 
$G(\zmax)$ for the open $w_0-w_a$ model; the forecast roughly 
follows these contours, but is tilted somewhat due to the 
$G_0(z)-G(\zmax)$ covariance from the CMB distance constraint.

To test how well the early dark energy parametrization with $\winf$ 
can model other forms of early dark energy evolution, we use the 
parametrization of Ref.~\cite{Doran06a} 
in which the dark energy fraction approaches a constant value 
$\Omega_e$ at high redshift:
\begin{equation}
\ode(a) = \frac{\ode - \Omega_e 
(1-a^{-3w_0})}{\ode+\om a^{3w_0}} 
+\Omega_e (1-a^{-3w_0}),
\label{eq:ededr}
\end{equation}
where $w_0 = w(z=0)$. (This form assumes $\ok=0$, so we only use it 
to simulate distance and growth data for a flat universe.)
The equation of state at high redshifts during 
matter domination is $w\approx 0$. However, since the transition between 
$w\approx w_0$ at low $z$ and $w\approx 0$ at high $z$ can occur 
at $z \ne \zmax$ the effective value of $\winf$ that best matches 
distance and growth observables for this model is not necessarily 
$\winf=0$.

We take the parameters for the simulated SNAP, Planck, and IXO data to be
$w_0=-0.9$, $\Omega_e=0.02$, $\om=0.225$, $h=0.73$, and $\ok=0$.
As for the previous early dark energy example, the constraint on 
$\ok$ marginalized over $\winf$ is biased due to the shape of the 
likelihood in the $\ok-\winf$ plane. However, for this model 
there is some additional bias such that the best fit is at 
$\ok\approx 0.004$. Given the increased width of the likelihood in $\ok$ 
[$\sigma(\ok)\sim 0.003$] and 
the long tail toward $\ok<0$, the true curvature $\ok=0$ is not 
strongly disfavored despite this bias.
The bias in $\ok$ is likely a reflection of the limitations of the $\winf$ 
parametrization of early dark energy, suggesting that a more 
flexible parametrization may be needed to accurately model a wide 
variety of early dark energy behavior. 

Additional priors can help reduce bias in curvature even 
with the default $\winf$ modeling of early dark energy in the 
growth reconstruction analysis.
The open models favored in the analysis of the Doran \& Robbers model 
are only able to fit the distance to recombination well by changing 
$\om$ and $H_0$ from their true values. Therefore, including 
a strong, $1 \%$ prior on $H_0$ in addition to the SN, CMB, and cluster 
data leads to nearly unbiased constraints on the curvature for 
this model of early dark energy.
Including the full CMB constraints instead of just the distance 
priors would also help in this case, since the $2\%$ early dark 
energy fraction at recombination would be detectable in Planck data.

For models with massive neutrinos, the results of the 
growth reconstruction analysis are similar to those for early 
dark energy models.
Massive neutrinos suppress growth on scales below their free streaming 
length \citep{Bond80a,Hu98a}, affecting the growth evolution in a 
manner similar to 
dark energy that tracks the dominant matter or radiation 
density at early times ($\winf=0$). 
Some of the strongest current 
cosmological limits on the neutrino mass come from the combination 
of the Lyman-$\alpha$ forest power spectrum with CMB data, 
with a 95\% CL upper limit on the sum of neutrino masses of 
$\sum m_{\nu}<0.17$~eV, assuming flat \lcdm\ \citep{Seljak06a}. 
More conservative upper bounds
limit the masses to $\sum m_{\nu}<1.3$~eV (95\% CL) using the CMB alone 
and $\sum m_{\nu}<0.67$~eV (95\% CL) with CMB, BAO, and SN data
(also assuming flat \lcdm\ \cite{Komatsu09a}). On the other hand, 
neutrino oscillation experiments indicate that there must be at least 
one neutrino mass eigenstate with $m_{\nu}\gtrsim 0.05$~eV 
(e.g., \cite{Michael06a,Schwetz08a}).

Using $\winf$ as an effective parameter to 
absorb the effects of massive neutrinos on distances and growth, 
the resulting curvature estimates tend to be biased toward 
high values of $\ok$.
The bias is similar to what we found for the Doran \& Robbers 
early dark energy model above, and as in that case an additional 
$H_0$ prior can reduce the curvature bias. 
Without any extra priors, the curvature can be biased by as much 
as $\sim0.5\sigma$ even for the minimal mass of $\sum m_{\nu}\approx 0.05$~eV,
so proper modeling of the effects of massive neutrinos is likely 
to be important for future curvature constraints from distances and growth.

Using the sum of neutrino masses as an additional parameter 
in the growth reconstruction analysis, instead of modeling the 
effects of neutrinos with $\winf$, produces unbiased 
curvature constraints. 
However, for models with massive neutrinos the uncertainty in 
curvature is fairly large 
due to a strong degeneracy between $\ok$ and $\sum m_{\nu}$.
In such a scenario, independent measurements of the neutrino masses 
would greatly reduce the uncertainty in curvature from distance and 
growth probes.
Future cosmological measurements such as weak lensing of the CMB 
may be able to determine neutrino masses with an accuracy of 
$\sigma(\sum m_{\nu})\approx 0.05$~eV, which should yield 
strong constraints on neutrino masses when combined with the 
results of terrestrial experiments (e.g., see \cite{Lesgourgues06a} for a review).

For the most general treatment of the high-redshift universe, 
one should ideally include parameters for both massive neutrinos 
and early dark energy models. We leave further study of this approach 
for future work, but note here that a simultaneous analysis of the 
impact of massive neutrinos and early dark energy on curvature estimation 
may be complicated by degeneracies between the two 
in the distance and growth observables.

%#####################################################
\subsection{Dependence on SN and CMB data modeling}
\label{sec:datadep1}

Using the growth reconstruction method,
we can study the relative contributions of scatter in the SN and CMB 
data to uncertainties in the reconstructed growth function by 
only including the scatter in one of these data sets in the 
Monte Carlo simulations. 
For the predicted growth observables $P_d({\bf g}|\ok)$,
the SN errors assumed for SNAP are the dominant source of uncertainty.
More precise CMB data than anticipated from Planck would have little 
impact on the uncertainties in the growth reconstruction.
However, for the \emph{curvature} constraint, more precise SN or CMB 
data would not significantly reduce $\sigma(\ok)$ since the 
assumed $1\%$ growth uncertainties for IXO clusters dominate the 
curvature uncertainty.

If we could obtain more precise growth measurements with uncertainties 
less than a few tenths of a percent, then the curvature estimate would 
be limited by the precision of the data from SNAP and Planck. However, the 
relative importance of these two data sets for curvature 
is opposite that for $P_d({\bf g}|\ok)$.
That is, given extremely precise growth measurements, $\sigma(\ok)$ 
would be reduced more by improving CMB data beyond Planck than by 
reducing SN uncertainties. The reason for this is related to the 
covariance between $G_0(z)$ and $G(\zmax)$: as explained above
(see Fig.~\ref{plot:okwinf}), CMB data 
induce a positive correlation between the reconstructions of  
$G_0(z)$ and $G(\zmax)$ from distance data. Since changing the 
curvature has opposite effects on $G_0(z)$ and $G(\zmax)$, the 
covariance from CMB constraints leads to more precise limits on 
curvature.
Improving CMB constraints therefore reduces curvature 
uncertainty by strengthening the $G_0(z)-G(\zmax)$ correlation.

The curvature forecasts are not strongly dependent on the 
redshift distribution of the supernova sample.  For example, 
if we assume a uniform distribution of 2300 SNe over $0<z<1.7$ 
instead of the SNAP distribution of 2000 SNe plus 300 low-$z$ SNe, 
the curvature estimate from SNAP+Planck+IXO is unchanged.

Interpolation of the distance--redshift relation between redshift bins 
or across gaps in the SN data introduces additional assumptions 
about the evolution of the expansion rate.  To test the impact of 
these assumptions, we have tried several distinct methods for 
interpolating between bins and found that the reconstructed growth 
changes by $\lesssim 1\%$, and the mean value of $\ok$ in 
forecasts changes by $\lesssim 5\times 10^{-4}$.
However, there is always an inherent assumption about smoothness 
of $\chi(z)$ in the growth reconstruction method that should 
be kept in mind when interpreting the results.

%#####################################################
\subsection{Dependence on cluster data modeling}
\label{sec:datadep2}

We have assumed fairly optimistic future growth constraints 
for the curvature forecasts, with $1\%$ measurements of the 
growth function at $z<1.5$ as anticipated for the proposed IXO cluster 
sample \citep{Vikhlinin09b}. We can easily study how the curvature 
forecasts depend on the assumed precision of the growth measurements
by rescaling the Fisher matrix of the growth observables
${\bf F}_g$ used in \eqn{pok2}. We find that for less optimistic assumptions 
about the growth uncertainties, it is still possible to obtain interesting 
model-independent estimates of the curvature. 
For example, doubling the growth uncertainties to $2\%$ 
increases the curvature uncertainty from $\sigma(\ok)=0.0022$ 
to $\sigma(\ok)=0.0033$, 
and for $3\%$ growth measurements, $\sigma(\ok)=0.0045$.

The forecasts presented in the previous sections assume that the 
cluster likelihood can be approximated as a multivariate Gaussian
distribution for uncorrelated growth observables $\{G_0(z_i)\}$ 
in $\Delta z=0.1$ redshift bins and $G(\zmax)$. In reality, we can 
expect that at least some of the growth observables will be 
correlated. Perhaps more importantly, the observed cluster abundance 
depends on distances and the expansion rate as well as the 
growth function, and there may be degeneracies between these 
functions in the cluster likelihood. We might 
expect that the effect on curvature estimates of treating
cluster abundance as purely a probe of the growth history
is small since the distance--redshift relation 
is well constrained by the SN and CMB data. 
Although a detailed study of curvature constraints using 
the full cluster likelihood is beyond the scope of this work, 
we present a simple test of this expectation here.

To check how well the fiducial distance data sets constrain 
degeneracies between distances and growth in the cluster 
likelihood, we compute the Fisher matrix for growth 
and distance 
assuming Poisson-distributed clusters \citep{Cash79a,Tegmark98a},
\begin{equation}
F_{\alpha\beta} = \sum_{i,j} \frac{1}{N(M_i,z_j)}\frac{\partial N(M_i,z_j)}{\partial \theta_{\alpha}}
\frac{\partial N(M_i,z_j)}{\partial \theta_{\beta}},
\label{eq:fishercl}
\end{equation}
%as $F_{ij} = -\partial^2 \ln \mathcal{L}/(\partial \theta_i \partial \theta_j)$,
where $\bm{\theta} = (\{G_0(z_j)\},\{H_0D(z_j)\})$, taking derivatives 
of the number of clusters in each bin $N(M_i,z_j)$ 
at the fiducial flat \lcdm\ model.
The mass bins $\{M_i\}$ have width $\Delta \ln M \approx 0.35$ and  
start at a mass threshold $M_{\rm min}$, and the redshift bins 
$\{z_j\}$ have width $\Delta z=0.1$ and cover the range $0.1\leq z\leq 1.5$.  
%We use a binned version of 
%the cluster likelihood from \cite{Vikhlinin09d}, 
%\begin{equation}
%-\ln \mathcal{L} = \sum_{i,j} \left[ N(M_i,z_j;\bm{\theta}) -
%N_{\rm fid}(M_i,z_j) \ln N(M_i,z_j;\bm{\theta}) \right],
%\label{eq:cllike}
%\end{equation}
%where $N_{\rm fid}$ is the number of clusters in each mass and redshift 
%bin in the fiducial cosmology, and $N(\bm{\theta})$ is the number 
%that would be expected in some other cosmology \citep{Cash79a}.

As described in Sec.~\ref{sec:grow}, the abundance of 
massive clusters is exponentially sensitive to growth through 
the mass function $dn/dM$ [\eqn{massfn}]. However, the observed number of 
clusters depends on the cosmological model in other ways as well.
In particular, the comoving volume element [\eqn{volume}] 
introduces additional dependence on the distance--redshift relation 
and expansion rate. The mass dependence of the effective volume in which 
a given survey probes clusters of a certain mass is also 
cosmology-dependent (and redshift-dependent) in general, although here for simplicity 
we take the volume to be constant above a mass threshold 
$M_{\rm min} = 10^{14}~h^{-1}~M_{\sun}$.

The main
mass proxy proposed for IXO is the product of the X-ray temperature $T_X$ 
and gas mass $M_{\rm gas}$, $Y_X=T_X M_{\rm gas}$, 
which is expected to have a relatively small, $<10\%$ scatter
based on cluster simulations \citep{Kravtsov06a}.
Assuming lognormal scatter in the $Y_X-M$ relation, the total 
number of clusters in redshift bin $z_j$ with width $\Delta z$ and 
mass bin $M_i<M<M_{i+1}$ is \citep{Vikhlinin09d}
\begin{eqnarray}
&&N(M_i,z_j)\approx \frac{\Delta z}{2} \int_{-\infty}^{\infty} d\ln M~
\frac{dn(M,z_j)}{d\ln M}~\frac{dV(z_j)}{d\Omega~dz} \\
&&\quad \times \left[\erf\left(\frac{\ln M_{i+1}-\ln M}{\sqrt{2}\sigma_{\ln M}}\right)
-\erf\left(\frac{\ln M_{i}-\ln M}{\sqrt{2}\sigma_{\ln M}}\right)\right].
\nonumber
\end{eqnarray}
We include a $3\%$ systematic error in $M$ in uncorrelated $\Delta z=0.1$ 
redshift bins to represent the IXO forecast for the error in the normalization of 
the $Y_X-M$ relation from weak lensing mass measurements of 100 clusters 
per redshift bin \citep{Vikhlinin09d}.

Cluster mass estimates generally also depend on 
the assumed cosmology, although the exact dependence varies for 
different mass proxies. 
Recent results from simulations indicate that this additional 
cosmological dependence is fairly weak \citep{Aghanim09a}, 
but it is nevertheless important to include in a full analysis of 
cluster data.
For simplicity, however, we neglect the 
cosmological dependence of the cluster masses here and assume that 
the sensitivity of cluster abundances to cosmology is 
dominated by the mass function and the volume element.

For the halo mass function [\eqn{massfn}], we use the parametrization fit to 
simulations in Ref.~\cite{Tinker08a},
\begin{equation}
f(\sigma) = A(z) \left[\left(\frac{\sigma}{b(z)}\right)^{-a(z)}+1\right] 
e^{-c/\sigma^2},\\
\end{equation}
where $A$, $a$, and $b$ are weakly redshift-dependent.
As the authors of that paper note, it is probably
more appropriate for the wide variety of cosmological models 
considered here to replace the redshift dependence of these 
parameters with dependence on the growth function.
In fact, it may be important to include dependence on not only 
the instantaneous value of the growth function at the 
redshift of a cluster, but also the evolution of the 
growth function prior to that redshift \citep{Ma07a}.
Recent studies using N-body simulations or semi-analytic modeling 
of nonlinear growth in cosmologies with more complex 
dark energy evolution than \lcdm\ have begun to examine 
such issues \citep{Linder05b,Bartelmann06a,Liberato06a,Basilakos07a,Francis07a,Francis09a,Francis09b,Grossi09a,Fedeli09a,Casarini09a}.
For the approximate modeling of the cluster mass function here,  
however, we use the simpler redshift dependence given 
by Ref.~\cite{Tinker08a}.

By taking the submatrix of $F_{\alpha\beta}$ from \eqn{fishercl} 
that corresponds to $\{G_0(z_j)\}$ only and inverting to get the 
covariance matrix, we find that
the growth uncertainties for \emph{fixed} distances are $\sim 1\%$.
Marginalizing over the distances by inverting the full Fisher matrix, 
including both growth and distance variables, increases the growth function 
uncertainties by a factor of $5-10$. However, adding SNAP-like 
SN distance constraints with $1\%$ accuracy in $\{H_0D(z_i)\}$
reduces the growth uncertainties after marginalizing over distances 
to $1-3\%$. Recall from the beginning of this section that 
3\% growth function uncertainties approximately double the error 
in the curvature estimate relative to 1\% growth uncertainties.

This test shows that distances at 
low $z$ should be constrained well enough by the future SN and CMB data that
the impact of additional cluster data can be roughly approximated by 
its expected accuracy on the growth function alone.
Including degeneracies between growth and distances in the halo 
mass function and comoving volume element results in somewhat 
weaker but still interesting constraints on curvature.
However, given the simplifying assumptions made here, 
curvature constraints from distance and growth data using 
the full cluster likelihood are a subject that deserves 
further exploration.

%#####################################################
\subsection{Comparison with other methods}
\label{sec:compare}

The model-independent curvature constraint from distances 
and growth has a forecasted accuracy of $\sigma(\ok)=0.002$ for 
SNAP SN, Planck CMB, and IXO cluster data, assuming that the 
true cosmology is close to flat \lcdm.
This is very close to the accuracy expected in the model-\emph{dependent}
context of \lcdm\ using the SN and CMB distance data only.
Thus the inclusion of growth information provides a curvature 
measurement that is free from possible biases due to assuming an
incorrect form of the dark energy evolution without sacrificing 
precision.

As we have seen, other types of measurements can play a similar role 
to the growth constraints. A $1\%$ measurement of the Hubble constant 
and a $1\%$ upper limit on the fraction of dark energy at recombination, 
when combined with SNAP SN and Planck CMB data, provide a 
model-independent curvature constraint with $\sigma(\ok)\sim 0.005$.
However, this constraint is strongly skewed with a long tail toward
open models, and without growth information the constraints such data 
can place on cosmologies that have significant amounts of 
early dark energy are severely limited.

The method of curvature estimation studied here can provide 
complementary constraints to other model-independent techniques.
Knox \cite{Knox06a} proposed using precise distance measurements, for example 
from BAO at high redshifts ($z\gtrsim 3$), in comparison with the 
distance to recombination measured in CMB data to probe curvature.
With future, percent-level BAO distances and Planck CMB data, 
this method is expected 
to attain an accuracy of $\sigma(\ok)\sim 0.001-0.002$ \citep{Knox06a,Knox06b,Hill08a}.
However, this measurement depends on the assumption that the 
universe is matter-dominated between the redshift of the BAO measurement 
and recombination. As a result, the estimated curvature is independent of the 
low-redshift dark energy modeling but still depends on the high-redshift 
dark energy evolution \citep{Knox06a}.  This dependence 
is similar to the degeneracy 
between curvature and early dark energy that we find when comparing 
high-$z$ SN distances to the CMB distance [\eqn{geomdegen2}].
Additional information, such as 
a measurement of $G(z)$ (or $\sigma_8$), or BAO in the line-of-sight direction to probe $H(z)$ at $z\sim 3$,
could help to reduce the high-$z$ model dependence of 
this method \citep{Knox06a}.

The curvature measurement proposed by Bernstein \cite{Bernstein06a} 
is perhaps the most model-independent method since it relies only on the 
FRW form of the metric without assuming particular dark energy 
properties and is valid for alternative theories of gravity as well.
Weak lensing galaxy--shear correlations can measure
distance ``triangles'' involving the lens distance, source distance, 
and the distance between the two, and the relations between 
these distances are sensitive to curvature. Future 
lensing and galaxy surveys can provide 
a purely geometric test of curvature with an expected accuracy of
$\sigma(\ok)\sim 0.02-0.04$ \citep{Bernstein06a,Zhan09a}. 
This technique is likely to be even less model-dependent than 
the distance plus growth method described here, but the forecasted 
uncertainties are an order of magnitude larger.

Given that each of these methods for obtaining model-independent 
curvature estimates relies on different types of data, it is 
difficult to compare forecasts directly. On the other hand, the 
existence of multiple methods using independent data sets 
means that there will be many opportunities for cross-checks of 
the curvature estimates. In this sense, it is useful to have an 
array of model-independent methods available that will have
different systematics, both theoretical and instrumental.

\vspace{.3cm}

%%%%%%%%%%%%%%%%%%%%%%%%%%%%%%%%%%%%%%%%%
\section{Discussion} \label{sec:conc}

With future supernova and CMB data sets making definite predictions 
for the growth of linear perturbations based on measured distances,
precise measurements of the 
actual growth history will provide model-independent estimates of 
spatial curvature. Such constraints have important implications for 
testing models of inflation, which predict $\ok\approx 0$, 
and for obtaining constraints on 
dark energy models that are robust to uncertainty in curvature.

If the true cosmology is similar to flat \lcdm, 
as current data would suggest, then a combination of SNe from SNAP, 
CMB data from Planck, and X-ray clusters from IXO can measure the 
curvature with an accuracy of $\sigma(\ok)\approx 0.002$, making 
minimal assumptions about the dark energy evolution.
The main constraint on curvature in this scenario comes from 
combining distance data with measurements of the evolution of the 
growth function at low redshifts. However, information about the 
normalization of growth relative to early times,
obtained by comparing cluster abundances with the amplitude of 
CMB anisotropies, can significantly reduce the uncertainty in curvature. 
The extra information 
mainly comes from the covariance between the growth normalization and 
the low-redshift growth evolution required by the CMB constraint 
on the distance to recombination.

These forecasts are robust to changes in the true value of the 
curvature and in the dark energy evolution at low redshift. 
However, if the true cosmology is significantly different from 
the concordance flat \lcdm\ model at high redshifts ($z\gtrsim 2$), 
then the forecasted errors on $\ok$ increase by a factor of a few 
and the constraints depend more strongly on the normalization of 
the low-redshift growth function relative to early times. 
Although model-independent limits on curvature alone are weaker 
in this scenario, the combination of future SN, CMB, and cluster data
would reduce the allowed model space to a narrow degeneracy between 
curvature and early dark energy or massive neutrinos. 
A precise independent measurement of the Hubble constant would 
mitigate possible biases in these constraints.

Estimates of curvature from distances and growth are complementary 
to other model-independent techniques. The projected accuracy is 
similar to that expected from comparing distances measured at $z\sim 3$ to the 
distance to recombination \citep{Knox06a}, but the inclusion of growth information 
reduces dependence on assumptions about the high-redshift
universe. Compared with proposed metric tests of curvature using weak lensing 
distance triangles \citep{Bernstein06a}, constraints from distances and 
growth are more model-dependent, in particular relying on the assumption 
that GR is valid on large scales and that dark energy clustering 
does not significantly affect the growth observables.  However, limits on 
curvature from combinations of distance and growth data are potentially 
much more precise.

We have focused here on the use of X-ray clusters to probe the 
growth history, but there are a number of other possible methods 
for measuring growth that may also provide interesting 
curvature constraints when combined with distance measurements.
Correlations involving weak lensing shear and galaxy density 
measurements should constrain both distances and growth in several 
redshift bins (e.g., \cite{Zhan09a}). Redshift space distortions of 
galaxy correlation functions can be used to constrain the growth rate
rather than the integrated growth, so such data sets may provide 
interesting complementary constraints to those from clusters or 
weak lensing. Forecasts of the ability of 
these alternative probes of growth to measure curvature when combined 
with distance data are an interesting subject for future study.

A fundamental assumption for using distances and growth to measure 
spatial curvature is that the relation between the growth history 
and the expansion rate is governed by GR. 
In the context of modified theories of gravity, many authors 
have studied the use of distances and growth to test for 
deviations from GR. These investigations typically 
make some simplifying assumptions regarding spatial flatness and/or 
the dark energy evolution, and relaxing these assumptions could 
make such tests of gravity considerably more complicated.
For example, at any particular scale the deviations in the 
distance--growth relation caused by modifying gravity may be 
difficult to distinguish from the effects of nonzero curvature or 
dynamical dark energy. However, scale dependence of growth 
could still be a robust signature of certain classes of modified gravity 
theories even in the context of more general cosmologies.
Regardless of our ability to use future data to 
distinguish between nonzero curvature, 
dynamical dark energy, and modified gravity, 
finding hints of any of these possibilities would be an 
intriguing sign of physics beyond the standard cosmological model.

\vspace{1cm}
{\it Acknowledgments:}
The author would like to thank Wayne Hu, Cora Dvorkin, and 
Amol Upadhye for useful discussions. This work was supported 
by the NSF GRFP.

\vspace{1cm}
\appendix

%%%%%%%%%%%%%%%%%%%%%%%%%%%%%%%%%%%%%%%%%%%%%%%%%%%%%%%%%
\section{Growth function approximation at high redshift}
\label{sec:growapprox}

At high redshift, assuming that the contributions of dark energy 
and curvature to $H(z)$ are much smaller than that of 
the matter density, we can derive an approximate form for the 
growth history. In an Einstein-de Sitter universe with matter only, 
the growth function is constant. We therefore start with an \emph{ansatz}
that the growth history in a universe with small fractions of 
dark energy and curvature is a small perturbation to 
the constant Einstein-de Sitter solution, expanding around 
some redshift $z_i$ where $G(z_i)=G_i$,
\begin{eqnarray}
G(z) = G_i &+& \cede\left[\left(\frac{1+z}{1+z_i}\right)^{3\winf}-1\right]
\nonumber \\
&+& \ccurv\left[\left(\frac{1+z}{1+z_i}\right)^{-1}-1\right],
\label{eq:earlygform}
\end{eqnarray}
where $\cede\ll G_i$, $\ccurv\ll G_i$, and we have assumed that dark energy 
at early times can be parametrized by a constant effective equation of 
state $\winf$. 
The redshift dependence of the terms in \eqn{earlygform} is the 
same as that of $\ode(z)$ and $\ok(z)$ during matter domination. 
The growth solution must have this dependence in order 
to solve \eqn{growth} 
since $d\ln H/d\ln a$ contains terms proportional to $\ode(z)$ and $\ok(z)$.

For this approximate solution to the growth equation, we neglect terms 
of second or higher order in $\cede$, $\ccurv$, $\ode(z_i)$, and 
$\ok(z_i)$. In this limit, the quantity $d\ln H/d\ln a$ appearing in 
\eqn{growth} is
\begin{equation}
\frac{d\ln H}{d\ln a} \approx -\frac{3}{2} + \frac{1}{2}\ok(z) -\frac{3}{2}
\winf\ode(z).
\label{eq:happrox}
\end{equation}
Using Eqs.~(\ref{eq:earlygform}) and~(\ref{eq:happrox}) 
in the differential equation for 
$G(z)$ produces algebraic relations for 
the early dark energy and curvature coefficients:
\begin{eqnarray}
\cede &=& \frac{1-\winf}{\winf(5-6\winf)}\ode(z_i)G_i,\\
\ccurv &=& -\frac{4}{7}\ok(z_i)G_i.
\end{eqnarray}

This approximation can be used to set the value of $\delta'(\zmax)$ 
needed for the analytic growth reconstruction in Sec.~\ref{sec:analytic}:
%\begin{eqnarray}
%\delta'(\zmax) &=& -\frac{\delta(\zmax)}{(1+\zmax)}\frac{H(\zmax)}{H_0}
%\left(1+\left.\frac{d\ln G}{d\ln a}\right|_{\zmax}\right),\label{eq:dpzmax}\\
%\left.\frac{d\ln G}{d\ln a}\right|_{\zmax} &=& -\frac{3(1-\winf)}{5-6\winf}\ode(\zmax)
%-\frac{4}{7}\ok(\zmax).\nonumber
%\end{eqnarray}
\begin{eqnarray}
\delta'(\zmax) &=& -\frac{\delta(\zmax)}{(1+\zmax)}\frac{H(\zmax)}{H_0}\nonumber\\
&&\times\left(1+\left.\frac{d\ln G}{d\ln a}\right|_{\zmax}\right),\label{eq:dpzmax}\\
\left.\frac{d\ln G}{d\ln a}\right|_{\zmax} &=& -\frac{3(1-\winf)}{5-6\winf}\ode(\zmax)
-\frac{4}{7}\ok(\zmax).\nonumber
\end{eqnarray}

The approximation breaks down when either the curvature or early 
dark energy fraction is large. For $\ode(\zmax)<0.25$ and $|\ok|<0.01$, 
the error in the approximation 
for $\delta'/\delta$ at $\zmax \approx 1.5$ is $\lesssim 1\%$. 
The approximate form of the growth rate in \eqn{dpzmax} remains accurate 
even as $\winf\to 0$ despite the fact that the integrated growth 
function [\eqn{earlygform}] becomes inaccurate at $\winf\gtrsim -1$.

\vspace{1cm}

%%%%%%%%%%%%%%%%%%%%%%%%%%%%%%%%%%%%%%%%%%%%%%%%%%%%%%%%%
\section{Monte Carlo simulations of growth reconstruction}
\label{sec:mcsims}

The procedure we use to estimate the covariance of 
growth reconstructed from distances (Sec.~\ref{sec:analytic}) 
for future SN and CMB data is as follows:
\begin{list}{\labelenumi}{\leftmargin=1em \usecounter{enumi} \listparindent=\parindent}
\item Define redshift bins $\{z_i\},~i=1,...,n_z$ with 
$z_1=0$ and $z_{n_{z}}=\zmax$.
\item Choose the assumed values of $\ok$ and early dark energy/massive neutrino 
parameters (here, we use $\winf$ as an example).
\item Draw a realization of the SN data, specified by the redshift 
distribution of Type Ia SNe and their magnitude errors: 
\begin{eqnarray}
\chi_\alpha &=& \frac{1}{\sqrt{|\ok|}}S_K^{-1}\left[\sqrt{|\ok|}(H_0D(z_\alpha)
+\epsilon_{\alpha})\right],\\
\epsilon_{\alpha}&=&\epsilon_{{\rm stat},\alpha} + \epsilon_{\rm sys}(z_{\alpha}),\nonumber
\end{eqnarray}
where the statistical error $\epsilon_{{\rm stat},\alpha}$ is
drawn for each SN (labeled by $\alpha$) from a Gaussian with width $\sigma_{H_0D,\alpha}=0.15 H_0D(z_\alpha)/(5\log e)$, 
and the systematic error $\epsilon_{\rm sys}(z_{\alpha})$ is drawn for each $\Delta z=0.1$ redshift bin 
from a Gaussian with width $0.02 [(1+z)/2.7]/(5\log e)$.
The SN distance--redshift relation is assumed to be unbiased relative 
to the fiducial model for the data, so $\langle \epsilon_{\alpha}\rangle = 0$.
\item Estimate $z(\chi)$ from the SN data by first estimating $\chi(z)$ 
and then inverting the relation. 
We compute $\chi(z)$ in redshift bins $z_i$ with width $\Delta z$ as
\begin{equation}
\chi(z_i) = \frac{\sum_\alpha \chi_\alpha \exp[-(z_\alpha-z_i)^2/(2\Delta z^2)]}
{\sum_\alpha \exp[-(z_\alpha-z_i)^2/(2\Delta z^2)]},
\label{eq:zest}
\end{equation}
where $z_{\alpha}$ and $\chi_{\alpha}$ are the redshifts and $\chi$ values 
of SNe from step 3. The default bin width used here is $\Delta z=0.05$.

To invert this to obtain $z(\chi)$, the estimated 
$\chi(z_i)$ must be monotonic. For large enough $\Delta z$ this is typically 
not a problem since the fiducial $\chi(z)$ relation is always 
monotonically increasing, 
but even with wide redshift bins there is a chance of having a few 
realizations with non-monotonic $\chi(z_i)$ estimates.
We correct for this when it 
occurs by simply setting $\chi(z_{i+1})=\chi(z_i)$ in any bin for 
which \eqn{zest} gives $\chi(z_{i+1})<\chi(z_i)$. The fact that the 
results are relatively independent of the redshift bin width indicates 
that this correction is not a significant source of error.

SN coverage that is fairly uniform in $z$ is important to avoid 
bias in the $z(\chi)$ relation. For the anticipated SNAP distribution 
the largest biases are at the ends, $z\approx 0$ and $z\approx \zmax$. 
The low-$z$ bias has little impact on the growth reconstruction at 
$z\gtrsim 0.1$, but the high-$z$ bias can cause problems with connecting 
the reconstructed growth at $z<\zmax$ to the fiducial $z>\zmax$ 
growth history. To avoid such problems, we take $\zmax$ for the growth 
reconstruction to be slightly smaller than the maximum SN redshift; 
in the case of SNAP where the SN distribution ends at $z=1.7$, $\zmax=1.5$
is sufficiently low to avoid problems relating to bias in $z(\chi)$.
(Note that using a different value of $\zmax$ results in 
slightly different definitions of early dark energy for the MCMC and 
growth reconstruction methods.)
\item Estimate $\emax = dz/d\chi(\zmax)$ using a linear fit to the 
high-$z$ end of the $z(\chi)$ relation.
\item Draw simulated CMB distance data from the $2\times 2$ Fisher matrix for 
$\{\ln(D_*/{\rm Mpc}),\omhh\}$ (Sec.~\ref{sec:data}).
Using the assumed curvature and early dark energy parameters 
in addition to the estimate of $\emax$ from step 5,
set $\om$ and $h$ to match the CMB constraints.
\item Optionally include additional priors (e.g.\ $H_0$ and dark 
energy fraction at last scattering).
Each Monte Carlo simulation is weighted by the likelihood associated 
with these priors, and the entire set of simulations for a single 
$\{\ok,\winf\}$ pair has a weight assigned to it equal to the 
average of the individual simulation weights.
\item Compute the growth at $\zmax$ relative to recombination, $G(\zmax)$, 
for the expansion history specified by $\ok$ and $\winf$ (step 2); 
$\emax$ (step 5); and $\om$ and $h$ (step 6).
\item Given $z(\chi)$ from step 4 and $\om$ from step 5, iteratively 
solve the growth reconstruction equation, \eqn{growrec2}.
For each iteration, $\dpmax$ is set to the average of its value in 
the previous iteration
and the target value based on the approximate growth evolution 
at $z>\zmax$ given by \eqn{dpzmax}.
\item Repeat steps $3-9$ for many realizations of the SN and CMB data, 
and compute the mean and covariance (using the weights from 
step 7) of the resulting estimates of 
${\bf g}=\{G(\zmax),G_0(z_i)\}$. This produces the predicted growth 
observables at fixed $\ok$ and $\winf$.
\item Repeat steps $2-10$ for different curvature and early dark energy 
parameter values to compute $P_d({\bf g}|\Omega_K,\winf)$.
\end{list}

\bibliography{thesis_mjm}

\begin{thebibliography}{89}
\expandafter\ifx\csname natexlab\endcsname\relax\def\natexlab#1{#1}\fi
\expandafter\ifx\csname bibnamefont\endcsname\relax
  \def\bibnamefont#1{#1}\fi
\expandafter\ifx\csname bibfnamefont\endcsname\relax
  \def\bibfnamefont#1{#1}\fi
\expandafter\ifx\csname citenamefont\endcsname\relax
  \def\citenamefont#1{#1}\fi
\expandafter\ifx\csname url\endcsname\relax
  \def\url#1{\texttt{#1}}\fi
\expandafter\ifx\csname urlprefix\endcsname\relax\def\urlprefix{URL }\fi
\providecommand{\bibinfo}[2]{#2}
\providecommand{\eprint}[2][]{\url{#2}}

\bibitem[{\citenamefont{{Komatsu} et~al.}(2009)\citenamefont{{Komatsu},
  {Dunkley}, {Nolta}, {Bennett}, {Gold}, {Hinshaw}, {Jarosik}, {Larson},
  {Limon}, {Page} et~al.}}]{Komatsu09a}
\bibinfo{author}{\bibfnamefont{E.}~\bibnamefont{{Komatsu}}},
  \bibinfo{author}{\bibfnamefont{J.}~\bibnamefont{{Dunkley}}},
  \bibinfo{author}{\bibfnamefont{M.~R.} \bibnamefont{{Nolta}}},
  \bibinfo{author}{\bibfnamefont{C.~L.} \bibnamefont{{Bennett}}},
  \bibinfo{author}{\bibfnamefont{B.}~\bibnamefont{{Gold}}},
  \bibinfo{author}{\bibfnamefont{G.}~\bibnamefont{{Hinshaw}}},
  \bibinfo{author}{\bibfnamefont{N.}~\bibnamefont{{Jarosik}}},
  \bibinfo{author}{\bibfnamefont{D.}~\bibnamefont{{Larson}}},
  \bibinfo{author}{\bibfnamefont{M.}~\bibnamefont{{Limon}}},
  \bibinfo{author}{\bibfnamefont{L.}~\bibnamefont{{Page}}},
  \bibnamefont{et~al.}, \bibinfo{journal}{\apjs}
  \textbf{\bibinfo{volume}{180}}, \bibinfo{pages}{330} (\bibinfo{year}{2009}),
  \eprint{arXiv:0803.0547}.

\bibitem[{\citenamefont{{Eisenstein} et~al.}(2005)\citenamefont{{Eisenstein},
  {Zehavi}, {Hogg}, {Scoccimarro}, {Blanton}, {Nichol}, {Scranton}, {Seo},
  {Tegmark}, {Zheng} et~al.}}]{Eisenstein05a}
\bibinfo{author}{\bibfnamefont{D.~J.} \bibnamefont{{Eisenstein}}},
  \bibinfo{author}{\bibfnamefont{I.}~\bibnamefont{{Zehavi}}},
  \bibinfo{author}{\bibfnamefont{D.~W.} \bibnamefont{{Hogg}}},
  \bibinfo{author}{\bibfnamefont{R.}~\bibnamefont{{Scoccimarro}}},
  \bibinfo{author}{\bibfnamefont{M.~R.} \bibnamefont{{Blanton}}},
  \bibinfo{author}{\bibfnamefont{R.~C.} \bibnamefont{{Nichol}}},
  \bibinfo{author}{\bibfnamefont{R.}~\bibnamefont{{Scranton}}},
  \bibinfo{author}{\bibfnamefont{H.-J.} \bibnamefont{{Seo}}},
  \bibinfo{author}{\bibfnamefont{M.}~\bibnamefont{{Tegmark}}},
  \bibinfo{author}{\bibfnamefont{Z.}~\bibnamefont{{Zheng}}},
  \bibnamefont{et~al.}, \bibinfo{journal}{\apj} \textbf{\bibinfo{volume}{633}},
  \bibinfo{pages}{560} (\bibinfo{year}{2005}), \eprint{arXiv:astro-ph/0501171}.

\bibitem[{\citenamefont{{Zhao} et~al.}(2007)\citenamefont{{Zhao}, {Xia}, {Li},
  {Tao}, {Virey}, {Zhu}, and {Zhang}}}]{Zhao07a}
\bibinfo{author}{\bibfnamefont{G.-B.} \bibnamefont{{Zhao}}},
  \bibinfo{author}{\bibfnamefont{J.-Q.} \bibnamefont{{Xia}}},
  \bibinfo{author}{\bibfnamefont{H.}~\bibnamefont{{Li}}},
  \bibinfo{author}{\bibfnamefont{C.}~\bibnamefont{{Tao}}},
  \bibinfo{author}{\bibfnamefont{J.-M.} \bibnamefont{{Virey}}},
  \bibinfo{author}{\bibfnamefont{Z.-H.} \bibnamefont{{Zhu}}}, \bibnamefont{and}
  \bibinfo{author}{\bibfnamefont{X.}~\bibnamefont{{Zhang}}},
  \bibinfo{journal}{Physics Letters B} \textbf{\bibinfo{volume}{648}},
  \bibinfo{pages}{8} (\bibinfo{year}{2007}), \eprint{arXiv:astro-ph/0612728}.

\bibitem[{\citenamefont{{Ichikawa} and {Takahashi}}(2007)}]{Ichikawa07a}
\bibinfo{author}{\bibfnamefont{K.}~\bibnamefont{{Ichikawa}}} \bibnamefont{and}
  \bibinfo{author}{\bibfnamefont{T.}~\bibnamefont{{Takahashi}}},
  \bibinfo{journal}{Journal of Cosmology and Astro-Particle Physics}
  \textbf{\bibinfo{volume}{2}}, \bibinfo{pages}{1} (\bibinfo{year}{2007}),
  \eprint{arXiv:astro-ph/0612739}.

\bibitem[{\citenamefont{{Wright}}(2007)}]{Wright07a}
\bibinfo{author}{\bibfnamefont{E.~L.} \bibnamefont{{Wright}}},
  \bibinfo{journal}{\apj} \textbf{\bibinfo{volume}{664}}, \bibinfo{pages}{633}
  (\bibinfo{year}{2007}), \eprint{arXiv:astro-ph/0701584}.

\bibitem[{\citenamefont{{Weinberg}}(1970)}]{Weinberg70a}
\bibinfo{author}{\bibfnamefont{S.}~\bibnamefont{{Weinberg}}},
  \bibinfo{journal}{\apjl} \textbf{\bibinfo{volume}{161}},
  \bibinfo{pages}{L233} (\bibinfo{year}{1970}).

\bibitem[{\citenamefont{{Bond} et~al.}(1997)\citenamefont{{Bond}, {Efstathiou},
  and {Tegmark}}}]{Bond97a}
\bibinfo{author}{\bibfnamefont{J.~R.} \bibnamefont{{Bond}}},
  \bibinfo{author}{\bibfnamefont{G.}~\bibnamefont{{Efstathiou}}},
  \bibnamefont{and}
  \bibinfo{author}{\bibfnamefont{M.}~\bibnamefont{{Tegmark}}},
  \bibinfo{journal}{\mnras} \textbf{\bibinfo{volume}{291}},
  \bibinfo{pages}{L33} (\bibinfo{year}{1997}), \eprint{arXiv:astro-ph/9702100}.

\bibitem[{\citenamefont{{Zaldarriaga} et~al.}(1997)\citenamefont{{Zaldarriaga},
  {Spergel}, and {Seljak}}}]{Zaldarriaga97a}
\bibinfo{author}{\bibfnamefont{M.}~\bibnamefont{{Zaldarriaga}}},
  \bibinfo{author}{\bibfnamefont{D.~N.} \bibnamefont{{Spergel}}},
  \bibnamefont{and} \bibinfo{author}{\bibfnamefont{U.}~\bibnamefont{{Seljak}}},
  \bibinfo{journal}{\apj} \textbf{\bibinfo{volume}{488}}, \bibinfo{pages}{1}
  (\bibinfo{year}{1997}), \eprint{arXiv:astro-ph/9702157}.

\bibitem[{\citenamefont{{Efstathiou} and {Bond}}(1999)}]{Efstathiou99a}
\bibinfo{author}{\bibfnamefont{G.}~\bibnamefont{{Efstathiou}}}
  \bibnamefont{and} \bibinfo{author}{\bibfnamefont{J.~R.}
  \bibnamefont{{Bond}}}, \bibinfo{journal}{\mnras}
  \textbf{\bibinfo{volume}{304}}, \bibinfo{pages}{75} (\bibinfo{year}{1999}),
  \eprint{arXiv:astro-ph/9807103}.

\bibitem[{\citenamefont{{Huey} et~al.}(1999)\citenamefont{{Huey}, {Wang},
  {Dave}, {Caldwell}, and {Steinhardt}}}]{Huey99a}
\bibinfo{author}{\bibfnamefont{G.}~\bibnamefont{{Huey}}},
  \bibinfo{author}{\bibfnamefont{L.}~\bibnamefont{{Wang}}},
  \bibinfo{author}{\bibfnamefont{R.}~\bibnamefont{{Dave}}},
  \bibinfo{author}{\bibfnamefont{R.~R.} \bibnamefont{{Caldwell}}},
  \bibnamefont{and} \bibinfo{author}{\bibfnamefont{P.~J.}
  \bibnamefont{{Steinhardt}}}, \bibinfo{journal}{\prd}
  \textbf{\bibinfo{volume}{59}}, \bibinfo{pages}{063005}
  (\bibinfo{year}{1999}), \eprint{arXiv:astro-ph/9804285}.

\bibitem[{\citenamefont{{Linder}}(2005)}]{Linder05a}
\bibinfo{author}{\bibfnamefont{E.~V.} \bibnamefont{{Linder}}},
  \bibinfo{journal}{Astroparticle Physics} \textbf{\bibinfo{volume}{24}},
  \bibinfo{pages}{391} (\bibinfo{year}{2005}), \eprint{arXiv:astro-ph/0508333}.

\bibitem[{\citenamefont{{Polarski} and {Ranquet}}(2005)}]{Polarski05a}
\bibinfo{author}{\bibfnamefont{D.}~\bibnamefont{{Polarski}}} \bibnamefont{and}
  \bibinfo{author}{\bibfnamefont{A.}~\bibnamefont{{Ranquet}}},
  \bibinfo{journal}{Physics Letters B} \textbf{\bibinfo{volume}{627}},
  \bibinfo{pages}{1} (\bibinfo{year}{2005}), \eprint{arXiv:astro-ph/0507290}.

\bibitem[{\citenamefont{{Clarkson} et~al.}(2007)\citenamefont{{Clarkson},
  {Cort{\^e}s}, and {Bassett}}}]{Clarkson07a}
\bibinfo{author}{\bibfnamefont{C.}~\bibnamefont{{Clarkson}}},
  \bibinfo{author}{\bibfnamefont{M.}~\bibnamefont{{Cort{\^e}s}}},
  \bibnamefont{and}
  \bibinfo{author}{\bibfnamefont{B.}~\bibnamefont{{Bassett}}},
  \bibinfo{journal}{Journal of Cosmology and Astro-Particle Physics}
  \textbf{\bibinfo{volume}{8}}, \bibinfo{pages}{11} (\bibinfo{year}{2007}),
  \eprint{arXiv:astro-ph/0702670}.

\bibitem[{\citenamefont{{Virey} et~al.}(2008)\citenamefont{{Virey},
  {Talon-Esmieu}, {Ealet}, {Taxil}, and {Tilquin}}}]{Virey08a}
\bibinfo{author}{\bibfnamefont{J.-M.} \bibnamefont{{Virey}}},
  \bibinfo{author}{\bibfnamefont{D.}~\bibnamefont{{Talon-Esmieu}}},
  \bibinfo{author}{\bibfnamefont{A.}~\bibnamefont{{Ealet}}},
  \bibinfo{author}{\bibfnamefont{P.}~\bibnamefont{{Taxil}}}, \bibnamefont{and}
  \bibinfo{author}{\bibfnamefont{A.}~\bibnamefont{{Tilquin}}},
  \bibinfo{journal}{Journal of Cosmology and Astro-Particle Physics}
  \textbf{\bibinfo{volume}{12}}, \bibinfo{pages}{8} (\bibinfo{year}{2008}),
  \eprint{arXiv:0802.4407}.

\bibitem[{\citenamefont{{Hlozek} et~al.}(2008)\citenamefont{{Hlozek},
  {Cort{\^e}s}, {Clarkson}, and {Bassett}}}]{Hlozek08a}
\bibinfo{author}{\bibfnamefont{R.}~\bibnamefont{{Hlozek}}},
  \bibinfo{author}{\bibfnamefont{M.}~\bibnamefont{{Cort{\^e}s}}},
  \bibinfo{author}{\bibfnamefont{C.}~\bibnamefont{{Clarkson}}},
  \bibnamefont{and}
  \bibinfo{author}{\bibfnamefont{B.}~\bibnamefont{{Bassett}}},
  \bibinfo{journal}{General Relativity and Gravitation}
  \textbf{\bibinfo{volume}{40}}, \bibinfo{pages}{285} (\bibinfo{year}{2008}),
  \eprint{arXiv:0801.3847}.

\bibitem[{\citenamefont{{Huang} et~al.}(2007)\citenamefont{{Huang}, {Wang}, and
  {Su}}}]{Huang07a}
\bibinfo{author}{\bibfnamefont{Z.-Y.} \bibnamefont{{Huang}}},
  \bibinfo{author}{\bibfnamefont{B.}~\bibnamefont{{Wang}}}, \bibnamefont{and}
  \bibinfo{author}{\bibfnamefont{R.-K.} \bibnamefont{{Su}}},
  \bibinfo{journal}{International Journal of Modern Physics A}
  \textbf{\bibinfo{volume}{22}}, \bibinfo{pages}{1819} (\bibinfo{year}{2007}),
  \eprint{arXiv:astro-ph/0605392}.

\bibitem[{\citenamefont{{Caldwell} and {Kamionkowski}}(2004)}]{Caldwell04a}
\bibinfo{author}{\bibfnamefont{R.~R.} \bibnamefont{{Caldwell}}}
  \bibnamefont{and}
  \bibinfo{author}{\bibfnamefont{M.}~\bibnamefont{{Kamionkowski}}},
  \bibinfo{journal}{Journal of Cosmology and Astro-Particle Physics}
  \textbf{\bibinfo{volume}{9}}, \bibinfo{pages}{9} (\bibinfo{year}{2004}),
  \eprint{arXiv:astro-ph/0403003}.

\bibitem[{\citenamefont{{Bernstein}}(2006)}]{Bernstein06a}
\bibinfo{author}{\bibfnamefont{G.}~\bibnamefont{{Bernstein}}},
  \bibinfo{journal}{\apj} \textbf{\bibinfo{volume}{637}}, \bibinfo{pages}{598}
  (\bibinfo{year}{2006}), \eprint{arXiv:astro-ph/0503276}.

\bibitem[{\citenamefont{{Knox}}(2006)}]{Knox06a}
\bibinfo{author}{\bibfnamefont{L.}~\bibnamefont{{Knox}}},
  \bibinfo{journal}{\prd} \textbf{\bibinfo{volume}{73}},
  \bibinfo{pages}{023503} (\bibinfo{year}{2006}),
  \eprint{arXiv:astro-ph/0503405}.

\bibitem[{\citenamefont{{Hu}}(2005)}]{Hu05a}
\bibinfo{author}{\bibfnamefont{W.}~\bibnamefont{{Hu}}}, in
  \emph{\bibinfo{booktitle}{Observing Dark Energy}}, edited by
  \bibinfo{editor}{\bibfnamefont{S.~C.} \bibnamefont{{Wolff}}}
  \bibnamefont{and} \bibinfo{editor}{\bibfnamefont{T.~R.}
  \bibnamefont{{Lauer}}} (\bibinfo{year}{2005}), vol. \bibinfo{volume}{339} of
  \emph{\bibinfo{series}{Astronomical Society of the Pacific Conference
  Series}}, p. \bibinfo{pages}{215}, \eprint{arXiv:astro-ph/0407158}.

\bibitem[{\citenamefont{{Hu} et~al.}(2006)\citenamefont{{Hu}, {Huterer}, and
  {Smith}}}]{Hu06b}
\bibinfo{author}{\bibfnamefont{W.}~\bibnamefont{{Hu}}},
  \bibinfo{author}{\bibfnamefont{D.}~\bibnamefont{{Huterer}}},
  \bibnamefont{and} \bibinfo{author}{\bibfnamefont{K.~M.}
  \bibnamefont{{Smith}}}, \bibinfo{journal}{\apjl}
  \textbf{\bibinfo{volume}{650}}, \bibinfo{pages}{L13} (\bibinfo{year}{2006}),
  \eprint{arXiv:astro-ph/0607316}.

\bibitem[{\citenamefont{{Mortonson} et~al.}(2009)\citenamefont{{Mortonson},
  {Hu}, and {Huterer}}}]{Mortonson09a}
\bibinfo{author}{\bibfnamefont{M.~J.} \bibnamefont{{Mortonson}}},
  \bibinfo{author}{\bibfnamefont{W.}~\bibnamefont{{Hu}}}, \bibnamefont{and}
  \bibinfo{author}{\bibfnamefont{D.}~\bibnamefont{{Huterer}}},
  \bibinfo{journal}{\prd} \textbf{\bibinfo{volume}{79}},
  \bibinfo{pages}{023004} (\bibinfo{year}{2009}), \eprint{arXiv:0810.1744}.

\bibitem[{\citenamefont{{Alam} et~al.}(2008)\citenamefont{{Alam}, {Sahni}, and
  {Starobinsky}}}]{Alam08a}
\bibinfo{author}{\bibfnamefont{U.}~\bibnamefont{{Alam}}},
  \bibinfo{author}{\bibfnamefont{V.}~\bibnamefont{{Sahni}}}, \bibnamefont{and}
  \bibinfo{author}{\bibfnamefont{A.~A.} \bibnamefont{{Starobinsky}}}
  (\bibinfo{year}{2008}), \eprint{arXiv:0812.2846}.

\bibitem[{\citenamefont{{Kashlinsky} et~al.}(1994)\citenamefont{{Kashlinsky},
  {Tkachev}, and {Frieman}}}]{Kashlinsky94a}
\bibinfo{author}{\bibfnamefont{A.}~\bibnamefont{{Kashlinsky}}},
  \bibinfo{author}{\bibfnamefont{I.~I.} \bibnamefont{{Tkachev}}},
  \bibnamefont{and}
  \bibinfo{author}{\bibfnamefont{J.}~\bibnamefont{{Frieman}}},
  \bibinfo{journal}{Physical Review Letters} \textbf{\bibinfo{volume}{73}},
  \bibinfo{pages}{1582} (\bibinfo{year}{1994}),
  \eprint{arXiv:astro-ph/9405024}.

\bibitem[{\citenamefont{{Bucher} et~al.}(1995)\citenamefont{{Bucher},
  {Goldhaber}, and {Turok}}}]{Bucher95a}
\bibinfo{author}{\bibfnamefont{M.}~\bibnamefont{{Bucher}}},
  \bibinfo{author}{\bibfnamefont{A.~S.} \bibnamefont{{Goldhaber}}},
  \bibnamefont{and} \bibinfo{author}{\bibfnamefont{N.}~\bibnamefont{{Turok}}},
  \bibinfo{journal}{\prd} \textbf{\bibinfo{volume}{52}}, \bibinfo{pages}{3314}
  (\bibinfo{year}{1995}), \eprint{arXiv:hep-ph/9411206}.

\bibitem[{\citenamefont{{Waterhouse} and {Zibin}}(2008)}]{Waterhouse08a}
\bibinfo{author}{\bibfnamefont{T.~P.} \bibnamefont{{Waterhouse}}}
  \bibnamefont{and} \bibinfo{author}{\bibfnamefont{J.~P.}
  \bibnamefont{{Zibin}}} (\bibinfo{year}{2008}), \eprint{arXiv:0804.1771}.

\bibitem[{\citenamefont{{Vardanyan} et~al.}(2009)\citenamefont{{Vardanyan},
  {Trotta}, and {Silk}}}]{Vardanyan09a}
\bibinfo{author}{\bibfnamefont{M.}~\bibnamefont{{Vardanyan}}},
  \bibinfo{author}{\bibfnamefont{R.}~\bibnamefont{{Trotta}}}, \bibnamefont{and}
  \bibinfo{author}{\bibfnamefont{J.}~\bibnamefont{{Silk}}},
  \bibinfo{journal}{\mnras} p. \bibinfo{pages}{812} (\bibinfo{year}{2009}),
  \eprint{arXiv:0901.3354}.

\bibitem[{\citenamefont{{Gott}}(1982)}]{Gott82a}
\bibinfo{author}{\bibfnamefont{J.~R.} \bibnamefont{{Gott}},
  \bibfnamefont{III}}, \bibinfo{journal}{\nat} \textbf{\bibinfo{volume}{295}},
  \bibinfo{pages}{304} (\bibinfo{year}{1982}).

\bibitem[{\citenamefont{{Ellis} et~al.}(1991)\citenamefont{{Ellis}, {Lyth}, and
  {Miji{\'c}}}}]{Ellis91a}
\bibinfo{author}{\bibfnamefont{G.~F.~R.} \bibnamefont{{Ellis}}},
  \bibinfo{author}{\bibfnamefont{D.~H.} \bibnamefont{{Lyth}}},
  \bibnamefont{and} \bibinfo{author}{\bibfnamefont{M.~B.}
  \bibnamefont{{Miji{\'c}}}}, \bibinfo{journal}{Physics Letters B}
  \textbf{\bibinfo{volume}{271}}, \bibinfo{pages}{52} (\bibinfo{year}{1991}).

\bibitem[{\citenamefont{{Sasaki} et~al.}(1993)\citenamefont{{Sasaki}, {Tanaka},
  {Yamamoto}, and {Yokoyama}}}]{Sasaki93a}
\bibinfo{author}{\bibfnamefont{M.}~\bibnamefont{{Sasaki}}},
  \bibinfo{author}{\bibfnamefont{T.}~\bibnamefont{{Tanaka}}},
  \bibinfo{author}{\bibfnamefont{K.}~\bibnamefont{{Yamamoto}}},
  \bibnamefont{and}
  \bibinfo{author}{\bibfnamefont{J.}~\bibnamefont{{Yokoyama}}},
  \bibinfo{journal}{Physics Letters B} \textbf{\bibinfo{volume}{317}},
  \bibinfo{pages}{510} (\bibinfo{year}{1993}).

\bibitem[{\citenamefont{{Linde} and {Mezhlumian}}(1995)}]{Linde95a}
\bibinfo{author}{\bibfnamefont{A.}~\bibnamefont{{Linde}}} \bibnamefont{and}
  \bibinfo{author}{\bibfnamefont{A.}~\bibnamefont{{Mezhlumian}}},
  \bibinfo{journal}{\prd} \textbf{\bibinfo{volume}{52}}, \bibinfo{pages}{6789}
  (\bibinfo{year}{1995}), \eprint{arXiv:astro-ph/9506017}.

\bibitem[{\citenamefont{{Hawking} and {Turok}}(1998)}]{Hawking98a}
\bibinfo{author}{\bibfnamefont{S.~W.} \bibnamefont{{Hawking}}}
  \bibnamefont{and} \bibinfo{author}{\bibfnamefont{N.}~\bibnamefont{{Turok}}},
  \bibinfo{journal}{Physics Letters B} \textbf{\bibinfo{volume}{425}},
  \bibinfo{pages}{25} (\bibinfo{year}{1998}), \eprint{arXiv:hep-th/9802030}.

\bibitem[{\citenamefont{{Gratton} et~al.}(2002)\citenamefont{{Gratton},
  {Lewis}, and {Turok}}}]{Gratton02a}
\bibinfo{author}{\bibfnamefont{S.}~\bibnamefont{{Gratton}}},
  \bibinfo{author}{\bibfnamefont{A.}~\bibnamefont{{Lewis}}}, \bibnamefont{and}
  \bibinfo{author}{\bibfnamefont{N.}~\bibnamefont{{Turok}}},
  \bibinfo{journal}{\prd} \textbf{\bibinfo{volume}{65}},
  \bibinfo{pages}{043513} (\bibinfo{year}{2002}),
  \eprint{arXiv:astro-ph/0111012}.

\bibitem[{\citenamefont{{Uzan} et~al.}(2003)\citenamefont{{Uzan}, {Kirchner},
  and {Ellis}}}]{Uzan03a}
\bibinfo{author}{\bibfnamefont{J.-P.} \bibnamefont{{Uzan}}},
  \bibinfo{author}{\bibfnamefont{U.}~\bibnamefont{{Kirchner}}},
  \bibnamefont{and} \bibinfo{author}{\bibfnamefont{G.~F.~R.}
  \bibnamefont{{Ellis}}}, \bibinfo{journal}{\mnras}
  \textbf{\bibinfo{volume}{344}}, \bibinfo{pages}{L65} (\bibinfo{year}{2003}),
  \eprint{arXiv:astro-ph/0302597}.

\bibitem[{\citenamefont{{Linde}}(2003)}]{Linde03a}
\bibinfo{author}{\bibfnamefont{A.}~\bibnamefont{{Linde}}},
  \bibinfo{journal}{Journal of Cosmology and Astro-Particle Physics}
  \textbf{\bibinfo{volume}{5}}, \bibinfo{pages}{2} (\bibinfo{year}{2003}),
  \eprint{arXiv:astro-ph/0303245}.

\bibitem[{\citenamefont{{Lasenby} and {Doran}}(2005)}]{Lasenby05a}
\bibinfo{author}{\bibfnamefont{A.}~\bibnamefont{{Lasenby}}} \bibnamefont{and}
  \bibinfo{author}{\bibfnamefont{C.}~\bibnamefont{{Doran}}},
  \bibinfo{journal}{\prd} \textbf{\bibinfo{volume}{71}},
  \bibinfo{pages}{063502} (\bibinfo{year}{2005}),
  \eprint{arXiv:astro-ph/0307311}.

\bibitem[{\citenamefont{{Freivogel} et~al.}(2006)\citenamefont{{Freivogel},
  {Kleban}, {Rodr{\'{\i}}guez Mart{\'{\i}}nez}, and {Susskind}}}]{Freivogel06a}
\bibinfo{author}{\bibfnamefont{B.}~\bibnamefont{{Freivogel}}},
  \bibinfo{author}{\bibfnamefont{M.}~\bibnamefont{{Kleban}}},
  \bibinfo{author}{\bibfnamefont{M.}~\bibnamefont{{Rodr{\'{\i}}guez
  Mart{\'{\i}}nez}}}, \bibnamefont{and}
  \bibinfo{author}{\bibfnamefont{L.}~\bibnamefont{{Susskind}}},
  \bibinfo{journal}{Journal of High Energy Physics}
  \textbf{\bibinfo{volume}{3}}, \bibinfo{pages}{39} (\bibinfo{year}{2006}),
  \eprint{arXiv:hep-th/0505232}.

\bibitem[{\citenamefont{{Spergel} et~al.}(2003)\citenamefont{{Spergel},
  {Verde}, {Peiris}, {Komatsu}, {Nolta}, {Bennett}, {Halpern}, {Hinshaw},
  {Jarosik}, {Kogut} et~al.}}]{Spergel03a}
\bibinfo{author}{\bibfnamefont{D.~N.} \bibnamefont{{Spergel}}},
  \bibinfo{author}{\bibfnamefont{L.}~\bibnamefont{{Verde}}},
  \bibinfo{author}{\bibfnamefont{H.~V.} \bibnamefont{{Peiris}}},
  \bibinfo{author}{\bibfnamefont{E.}~\bibnamefont{{Komatsu}}},
  \bibinfo{author}{\bibfnamefont{M.~R.} \bibnamefont{{Nolta}}},
  \bibinfo{author}{\bibfnamefont{C.~L.} \bibnamefont{{Bennett}}},
  \bibinfo{author}{\bibfnamefont{M.}~\bibnamefont{{Halpern}}},
  \bibinfo{author}{\bibfnamefont{G.}~\bibnamefont{{Hinshaw}}},
  \bibinfo{author}{\bibfnamefont{N.}~\bibnamefont{{Jarosik}}},
  \bibinfo{author}{\bibfnamefont{A.}~\bibnamefont{{Kogut}}},
  \bibnamefont{et~al.}, \bibinfo{journal}{\apjs}
  \textbf{\bibinfo{volume}{148}}, \bibinfo{pages}{175} (\bibinfo{year}{2003}),
  \eprint{arXiv:astro-ph/0302209}.

\bibitem[{\citenamefont{{Rozo} et~al.}(2009)\citenamefont{{Rozo}, {Wechsler},
  {Rykoff}, {Annis}, {Becker}, {Evrard}, {Frieman}, {Hansen}, {Hao}, {Johnston}
  et~al.}}]{Rozo09a}
\bibinfo{author}{\bibfnamefont{E.}~\bibnamefont{{Rozo}}},
  \bibinfo{author}{\bibfnamefont{R.~H.} \bibnamefont{{Wechsler}}},
  \bibinfo{author}{\bibfnamefont{E.~S.} \bibnamefont{{Rykoff}}},
  \bibinfo{author}{\bibfnamefont{J.~T.} \bibnamefont{{Annis}}},
  \bibinfo{author}{\bibfnamefont{M.~R.} \bibnamefont{{Becker}}},
  \bibinfo{author}{\bibfnamefont{A.~E.} \bibnamefont{{Evrard}}},
  \bibinfo{author}{\bibfnamefont{J.~A.} \bibnamefont{{Frieman}}},
  \bibinfo{author}{\bibfnamefont{S.~M.} \bibnamefont{{Hansen}}},
  \bibinfo{author}{\bibfnamefont{J.}~\bibnamefont{{Hao}}},
  \bibinfo{author}{\bibfnamefont{D.~E.} \bibnamefont{{Johnston}}},
  \bibnamefont{et~al.} (\bibinfo{year}{2009}), \eprint{arXiv:0902.3702}.

\bibitem[{\citenamefont{{Vikhlinin}
  et~al.}(2009{\natexlab{a}})\citenamefont{{Vikhlinin}, {Kravtsov}, {Burenin},
  {Ebeling}, {Forman}, {Hornstrup}, {Jones}, {Murray}, {Nagai}, {Quintana}
  et~al.}}]{Vikhlinin09a}
\bibinfo{author}{\bibfnamefont{A.}~\bibnamefont{{Vikhlinin}}},
  \bibinfo{author}{\bibfnamefont{A.~V.} \bibnamefont{{Kravtsov}}},
  \bibinfo{author}{\bibfnamefont{R.~A.} \bibnamefont{{Burenin}}},
  \bibinfo{author}{\bibfnamefont{H.}~\bibnamefont{{Ebeling}}},
  \bibinfo{author}{\bibfnamefont{W.~R.} \bibnamefont{{Forman}}},
  \bibinfo{author}{\bibfnamefont{A.}~\bibnamefont{{Hornstrup}}},
  \bibinfo{author}{\bibfnamefont{C.}~\bibnamefont{{Jones}}},
  \bibinfo{author}{\bibfnamefont{S.~S.} \bibnamefont{{Murray}}},
  \bibinfo{author}{\bibfnamefont{D.}~\bibnamefont{{Nagai}}},
  \bibinfo{author}{\bibfnamefont{H.}~\bibnamefont{{Quintana}}},
  \bibnamefont{et~al.}, \bibinfo{journal}{\apj} \textbf{\bibinfo{volume}{692}},
  \bibinfo{pages}{1060} (\bibinfo{year}{2009}{\natexlab{a}}),
  \eprint{arXiv:0812.2720}.

\bibitem[{\citenamefont{Ratra and Peebles}(1988)}]{Ratra88a}
\bibinfo{author}{\bibfnamefont{B.}~\bibnamefont{Ratra}} \bibnamefont{and}
  \bibinfo{author}{\bibfnamefont{P.~J.~E.} \bibnamefont{Peebles}},
  \bibinfo{journal}{\prd} \textbf{\bibinfo{volume}{37}}, \bibinfo{pages}{3406}
  (\bibinfo{year}{1988}).

\bibitem[{\citenamefont{Ferreira and Joyce}(1998)}]{Ferreira97a}
\bibinfo{author}{\bibfnamefont{P.~G.} \bibnamefont{Ferreira}} \bibnamefont{and}
  \bibinfo{author}{\bibfnamefont{M.}~\bibnamefont{Joyce}},
  \bibinfo{journal}{\prd} \textbf{\bibinfo{volume}{58}},
  \bibinfo{pages}{023503} (\bibinfo{year}{1998}),
  \eprint{arXiv:astro-ph/9711102}.

\bibitem[{\citenamefont{Steinhardt et~al.}(1999)\citenamefont{Steinhardt, Wang,
  and Zlatev}}]{Steinhardt98a}
\bibinfo{author}{\bibfnamefont{P.~J.} \bibnamefont{Steinhardt}},
  \bibinfo{author}{\bibfnamefont{L.-M.} \bibnamefont{Wang}}, \bibnamefont{and}
  \bibinfo{author}{\bibfnamefont{I.}~\bibnamefont{Zlatev}},
  \bibinfo{journal}{\prd} \textbf{\bibinfo{volume}{59}},
  \bibinfo{pages}{123504} (\bibinfo{year}{1999}),
  \eprint{arXiv:astro-ph/9812313}.

\bibitem[{\citenamefont{Aldering et~al.}(2004)}]{SNAP}
\bibinfo{author}{\bibfnamefont{G.}~\bibnamefont{Aldering}} \bibnamefont{et~al.}
  (\bibinfo{year}{2004}), \eprint{arXiv:astro-ph/0405232}.

\bibitem[{\citenamefont{{The Planck Collaboration}}(2006)}]{Planck}
\bibinfo{author}{\bibnamefont{{The Planck Collaboration}}}
  (\bibinfo{year}{2006}), \eprint{arXiv:astro-ph/0604069}.

\bibitem[{\citenamefont{{Freedman} et~al.}(2001)\citenamefont{{Freedman},
  {Madore}, {Gibson}, {Ferrarese}, {Kelson}, {Sakai}, {Mould}, {Kennicutt},
  {Ford}, {Graham} et~al.}}]{Freedman01a}
\bibinfo{author}{\bibfnamefont{W.~L.} \bibnamefont{{Freedman}}},
  \bibinfo{author}{\bibfnamefont{B.~F.} \bibnamefont{{Madore}}},
  \bibinfo{author}{\bibfnamefont{B.~K.} \bibnamefont{{Gibson}}},
  \bibinfo{author}{\bibfnamefont{L.}~\bibnamefont{{Ferrarese}}},
  \bibinfo{author}{\bibfnamefont{D.~D.} \bibnamefont{{Kelson}}},
  \bibinfo{author}{\bibfnamefont{S.}~\bibnamefont{{Sakai}}},
  \bibinfo{author}{\bibfnamefont{J.~R.} \bibnamefont{{Mould}}},
  \bibinfo{author}{\bibfnamefont{R.~C.} \bibnamefont{{Kennicutt}},
  \bibfnamefont{Jr.}}, \bibinfo{author}{\bibfnamefont{H.~C.}
  \bibnamefont{{Ford}}}, \bibinfo{author}{\bibfnamefont{J.~A.}
  \bibnamefont{{Graham}}}, \bibnamefont{et~al.}, \bibinfo{journal}{\apj}
  \textbf{\bibinfo{volume}{553}}, \bibinfo{pages}{47} (\bibinfo{year}{2001}),
  \eprint{arXiv:astro-ph/0012376}.

\bibitem[{\citenamefont{{Doran} et~al.}(2007)\citenamefont{{Doran}, {Robbers},
  and {Wetterich}}}]{Doran07a}
\bibinfo{author}{\bibfnamefont{M.}~\bibnamefont{{Doran}}},
  \bibinfo{author}{\bibfnamefont{G.}~\bibnamefont{{Robbers}}},
  \bibnamefont{and}
  \bibinfo{author}{\bibfnamefont{C.}~\bibnamefont{{Wetterich}}},
  \bibinfo{journal}{\prd} \textbf{\bibinfo{volume}{75}},
  \bibinfo{pages}{023003} (\bibinfo{year}{2007}),
  \eprint{arXiv:astro-ph/0609814}.

\bibitem[{\citenamefont{Greenhill et~al.}(2009)}]{Greenhill09a}
\bibinfo{author}{\bibfnamefont{L.}~\bibnamefont{Greenhill}}
  \bibnamefont{et~al.} (\bibinfo{year}{2009}), \eprint{arXiv:0902.4255}.

\bibitem[{\citenamefont{{de Putter} et~al.}(2009)\citenamefont{{de Putter},
  {Zahn}, and {Linder}}}]{dePutter09a}
\bibinfo{author}{\bibfnamefont{R.}~\bibnamefont{{de Putter}}},
  \bibinfo{author}{\bibfnamefont{O.}~\bibnamefont{{Zahn}}}, \bibnamefont{and}
  \bibinfo{author}{\bibfnamefont{E.~V.} \bibnamefont{{Linder}}},
  \bibinfo{journal}{\prd} \textbf{\bibinfo{volume}{79}},
  \bibinfo{pages}{065033} (\bibinfo{year}{2009}), \eprint{arXiv:0901.0916}.

\bibitem[{\citenamefont{{Hollenstein} et~al.}(2009)\citenamefont{{Hollenstein},
  {Sapone}, {Crittenden}, and {Sch{\"a}fer}}}]{Hollenstein09a}
\bibinfo{author}{\bibfnamefont{L.}~\bibnamefont{{Hollenstein}}},
  \bibinfo{author}{\bibfnamefont{D.}~\bibnamefont{{Sapone}}},
  \bibinfo{author}{\bibfnamefont{R.}~\bibnamefont{{Crittenden}}},
  \bibnamefont{and} \bibinfo{author}{\bibfnamefont{B.~M.}
  \bibnamefont{{Sch{\"a}fer}}}, \bibinfo{journal}{Journal of Cosmology and
  Astro-Particle Physics} \textbf{\bibinfo{volume}{4}}, \bibinfo{pages}{12}
  (\bibinfo{year}{2009}), \eprint{arXiv:0902.1494}.

\bibitem[{\citenamefont{{Vikhlinin}
  et~al.}(2009{\natexlab{b}})\citenamefont{{Vikhlinin}, {Allen}, {Arnaud},
  {Bautz}, {Boehringer}, {Bonamente}, {Burns}, {Evrard}, {Henry}, {Jones}
  et~al.}}]{Vikhlinin09b}
\bibinfo{author}{\bibfnamefont{A.}~\bibnamefont{{Vikhlinin}}},
  \bibinfo{author}{\bibfnamefont{S.~W.} \bibnamefont{{Allen}}},
  \bibinfo{author}{\bibfnamefont{M.}~\bibnamefont{{Arnaud}}},
  \bibinfo{author}{\bibfnamefont{M.}~\bibnamefont{{Bautz}}},
  \bibinfo{author}{\bibfnamefont{H.}~\bibnamefont{{Boehringer}}},
  \bibinfo{author}{\bibfnamefont{M.}~\bibnamefont{{Bonamente}}},
  \bibinfo{author}{\bibfnamefont{J.}~\bibnamefont{{Burns}}},
  \bibinfo{author}{\bibfnamefont{A.}~\bibnamefont{{Evrard}}},
  \bibinfo{author}{\bibfnamefont{J.~P.} \bibnamefont{{Henry}}},
  \bibinfo{author}{\bibfnamefont{C.}~\bibnamefont{{Jones}}},
  \bibnamefont{et~al.} (\bibinfo{year}{2009}{\natexlab{b}}),
  \eprint{arXiv:0903.2297}.

\bibitem[{\citenamefont{{Zahn} and {Zaldarriaga}}(2003)}]{Zahn03a}
\bibinfo{author}{\bibfnamefont{O.}~\bibnamefont{{Zahn}}} \bibnamefont{and}
  \bibinfo{author}{\bibfnamefont{M.}~\bibnamefont{{Zaldarriaga}}},
  \bibinfo{journal}{\prd} \textbf{\bibinfo{volume}{67}},
  \bibinfo{pages}{063002} (\bibinfo{year}{2003}),
  \eprint{arXiv:astro-ph/0212360}.

\bibitem[{\citenamefont{{Mortonson} and {Hu}}(2008)}]{Mortonson08a}
\bibinfo{author}{\bibfnamefont{M.~J.} \bibnamefont{{Mortonson}}}
  \bibnamefont{and} \bibinfo{author}{\bibfnamefont{W.}~\bibnamefont{{Hu}}},
  \bibinfo{journal}{\apj} \textbf{\bibinfo{volume}{672}}, \bibinfo{pages}{737}
  (\bibinfo{year}{2008}), \eprint{arXiv:0705.1132}.

\bibitem[{\citenamefont{Christensen et~al.}(2001)\citenamefont{Christensen,
  Meyer, Knox, and Luey}}]{Christensen01a}
\bibinfo{author}{\bibfnamefont{N.}~\bibnamefont{Christensen}},
  \bibinfo{author}{\bibfnamefont{R.}~\bibnamefont{Meyer}},
  \bibinfo{author}{\bibfnamefont{L.}~\bibnamefont{Knox}}, \bibnamefont{and}
  \bibinfo{author}{\bibfnamefont{B.}~\bibnamefont{Luey}},
  \bibinfo{journal}{Class. Quant. Grav.} \textbf{\bibinfo{volume}{18}},
  \bibinfo{pages}{2677} (\bibinfo{year}{2001}),
  \eprint{arXiv:astro-ph/0103134}.

\bibitem[{\citenamefont{Kosowsky et~al.}(2002)\citenamefont{Kosowsky,
  Milosavljevic, and Jimenez}}]{Kosowsky02a}
\bibinfo{author}{\bibfnamefont{A.}~\bibnamefont{Kosowsky}},
  \bibinfo{author}{\bibfnamefont{M.}~\bibnamefont{Milosavljevic}},
  \bibnamefont{and} \bibinfo{author}{\bibfnamefont{R.}~\bibnamefont{Jimenez}},
  \bibinfo{journal}{\prd} \textbf{\bibinfo{volume}{66}},
  \bibinfo{pages}{063007} (\bibinfo{year}{2002}),
  \eprint{arXiv:astro-ph/0206014}.

\bibitem[{\citenamefont{{Dunkley} et~al.}(2005)\citenamefont{{Dunkley},
  {Bucher}, {Ferreira}, {Moodley}, and {Skordis}}}]{Dunkley05a}
\bibinfo{author}{\bibfnamefont{J.}~\bibnamefont{{Dunkley}}},
  \bibinfo{author}{\bibfnamefont{M.}~\bibnamefont{{Bucher}}},
  \bibinfo{author}{\bibfnamefont{P.~G.} \bibnamefont{{Ferreira}}},
  \bibinfo{author}{\bibfnamefont{K.}~\bibnamefont{{Moodley}}},
  \bibnamefont{and}
  \bibinfo{author}{\bibfnamefont{C.}~\bibnamefont{{Skordis}}},
  \bibinfo{journal}{\mnras} \textbf{\bibinfo{volume}{356}},
  \bibinfo{pages}{925} (\bibinfo{year}{2005}), \eprint{arXiv:astro-ph/0405462}.

\bibitem[{\citenamefont{Gelman and Rubin}(1992)}]{Gelman92a}
\bibinfo{author}{\bibfnamefont{A.}~\bibnamefont{Gelman}} \bibnamefont{and}
  \bibinfo{author}{\bibfnamefont{D.}~\bibnamefont{Rubin}},
  \bibinfo{journal}{Statistical Science} \textbf{\bibinfo{volume}{7}},
  \bibinfo{pages}{452} (\bibinfo{year}{1992}).

\bibitem[{\citenamefont{{Upadhye} et~al.}(2005)\citenamefont{{Upadhye},
  {Ishak}, and {Steinhardt}}}]{Upadhye05a}
\bibinfo{author}{\bibfnamefont{A.}~\bibnamefont{{Upadhye}}},
  \bibinfo{author}{\bibfnamefont{M.}~\bibnamefont{{Ishak}}}, \bibnamefont{and}
  \bibinfo{author}{\bibfnamefont{P.~J.} \bibnamefont{{Steinhardt}}},
  \bibinfo{journal}{\prd} \textbf{\bibinfo{volume}{72}},
  \bibinfo{pages}{063501} (\bibinfo{year}{2005}),
  \eprint{arXiv:astro-ph/0411803}.

\bibitem[{\citenamefont{{Lewis} and {Bridle}}(2002)}]{Lewis02a}
\bibinfo{author}{\bibfnamefont{A.}~\bibnamefont{{Lewis}}} \bibnamefont{and}
  \bibinfo{author}{\bibfnamefont{S.}~\bibnamefont{{Bridle}}},
  \bibinfo{journal}{\prd} \textbf{\bibinfo{volume}{66}},
  \bibinfo{pages}{103511} (\bibinfo{year}{2002}),
  \eprint{arXiv:astro-ph/0205436}.

\bibitem[{\citenamefont{{Sahni} and {Starobinsky}}(2006)}]{Sahni06a}
\bibinfo{author}{\bibfnamefont{V.}~\bibnamefont{{Sahni}}} \bibnamefont{and}
  \bibinfo{author}{\bibfnamefont{A.}~\bibnamefont{{Starobinsky}}},
  \bibinfo{journal}{International Journal of Modern Physics D}
  \textbf{\bibinfo{volume}{15}}, \bibinfo{pages}{2105} (\bibinfo{year}{2006}),
  \eprint{arXiv:astro-ph/0610026}.

\bibitem[{\citenamefont{{Chevallier} and {Polarski}}(2001)}]{Chevallier01a}
\bibinfo{author}{\bibfnamefont{M.}~\bibnamefont{{Chevallier}}}
  \bibnamefont{and}
  \bibinfo{author}{\bibfnamefont{D.}~\bibnamefont{{Polarski}}},
  \bibinfo{journal}{International Journal of Modern Physics D}
  \textbf{\bibinfo{volume}{10}}, \bibinfo{pages}{213} (\bibinfo{year}{2001}),
  \eprint{arXiv:gr-qc/0009008}.

\bibitem[{\citenamefont{{Linder}}(2003)}]{Linder03a}
\bibinfo{author}{\bibfnamefont{E.~V.} \bibnamefont{{Linder}}},
  \bibinfo{journal}{Physical Review Letters} \textbf{\bibinfo{volume}{90}},
  \bibinfo{pages}{091301} (\bibinfo{year}{2003}),
  \eprint{arXiv:astro-ph/0208512}.

\bibitem[{\citenamefont{{Doran} and {Robbers}}(2006)}]{Doran06a}
\bibinfo{author}{\bibfnamefont{M.}~\bibnamefont{{Doran}}} \bibnamefont{and}
  \bibinfo{author}{\bibfnamefont{G.}~\bibnamefont{{Robbers}}},
  \bibinfo{journal}{Journal of Cosmology and Astro-Particle Physics}
  \textbf{\bibinfo{volume}{6}}, \bibinfo{pages}{26} (\bibinfo{year}{2006}),
  \eprint{arXiv:astro-ph/0601544}.

\bibitem[{\citenamefont{{Bond} et~al.}(1980)\citenamefont{{Bond}, {Efstathiou},
  and {Silk}}}]{Bond80a}
\bibinfo{author}{\bibfnamefont{J.~R.} \bibnamefont{{Bond}}},
  \bibinfo{author}{\bibfnamefont{G.}~\bibnamefont{{Efstathiou}}},
  \bibnamefont{and} \bibinfo{author}{\bibfnamefont{J.}~\bibnamefont{{Silk}}},
  \bibinfo{journal}{Physical Review Letters} \textbf{\bibinfo{volume}{45}},
  \bibinfo{pages}{1980} (\bibinfo{year}{1980}).

\bibitem[{\citenamefont{{Hu} and {Eisenstein}}(1998)}]{Hu98a}
\bibinfo{author}{\bibfnamefont{W.}~\bibnamefont{{Hu}}} \bibnamefont{and}
  \bibinfo{author}{\bibfnamefont{D.~J.} \bibnamefont{{Eisenstein}}},
  \bibinfo{journal}{\apj} \textbf{\bibinfo{volume}{498}}, \bibinfo{pages}{497}
  (\bibinfo{year}{1998}), \eprint{arXiv:astro-ph/9710216}.

\bibitem[{\citenamefont{{Seljak} et~al.}(2006)\citenamefont{{Seljak}, {Slosar},
  and {McDonald}}}]{Seljak06a}
\bibinfo{author}{\bibfnamefont{U.}~\bibnamefont{{Seljak}}},
  \bibinfo{author}{\bibfnamefont{A.}~\bibnamefont{{Slosar}}}, \bibnamefont{and}
  \bibinfo{author}{\bibfnamefont{P.}~\bibnamefont{{McDonald}}},
  \bibinfo{journal}{Journal of Cosmology and Astro-Particle Physics}
  \textbf{\bibinfo{volume}{10}}, \bibinfo{pages}{14} (\bibinfo{year}{2006}),
  \eprint{arXiv:astro-ph/0604335}.

\bibitem[{\citenamefont{{Michael} et~al.}(2006)}]{Michael06a}
\bibinfo{author}{\bibfnamefont{D.~G.} \bibnamefont{{Michael}}}
  \bibnamefont{et~al.}, \bibinfo{journal}{Physical Review Letters}
  \textbf{\bibinfo{volume}{97}}, \bibinfo{pages}{191801}
  (\bibinfo{year}{2006}), \eprint{arXiv:hep-ex/0607088}.

\bibitem[{\citenamefont{{Schwetz} et~al.}(2008)\citenamefont{{Schwetz},
  {T{\'o}rtola}, and {Valle}}}]{Schwetz08a}
\bibinfo{author}{\bibfnamefont{T.}~\bibnamefont{{Schwetz}}},
  \bibinfo{author}{\bibfnamefont{M.}~\bibnamefont{{T{\'o}rtola}}},
  \bibnamefont{and} \bibinfo{author}{\bibfnamefont{J.~W.~F.}
  \bibnamefont{{Valle}}}, \bibinfo{journal}{New Journal of Physics}
  \textbf{\bibinfo{volume}{10}}, \bibinfo{pages}{113011}
  (\bibinfo{year}{2008}), \eprint{arXiv:0808.2016}.

\bibitem[{\citenamefont{{Lesgourgues} and {Pastor}}(2006)}]{Lesgourgues06a}
\bibinfo{author}{\bibfnamefont{J.}~\bibnamefont{{Lesgourgues}}}
  \bibnamefont{and} \bibinfo{author}{\bibfnamefont{S.}~\bibnamefont{{Pastor}}},
  \bibinfo{journal}{\physrep} \textbf{\bibinfo{volume}{429}},
  \bibinfo{pages}{307} (\bibinfo{year}{2006}), \eprint{arXiv:astro-ph/0603494}.

\bibitem[{\citenamefont{{Cash}}(1979)}]{Cash79a}
\bibinfo{author}{\bibfnamefont{W.}~\bibnamefont{{Cash}}},
  \bibinfo{journal}{\apj} \textbf{\bibinfo{volume}{228}}, \bibinfo{pages}{939}
  (\bibinfo{year}{1979}).

\bibitem[{\citenamefont{{Tegmark} et~al.}(1998)\citenamefont{{Tegmark},
  {Eisenstein}, {Hu}, and {Kron}}}]{Tegmark98a}
\bibinfo{author}{\bibfnamefont{M.}~\bibnamefont{{Tegmark}}},
  \bibinfo{author}{\bibfnamefont{D.~J.} \bibnamefont{{Eisenstein}}},
  \bibinfo{author}{\bibfnamefont{W.}~\bibnamefont{{Hu}}}, \bibnamefont{and}
  \bibinfo{author}{\bibfnamefont{R.}~\bibnamefont{{Kron}}}
  (\bibinfo{year}{1998}), \eprint{arXiv:astro-ph/9805117}.

\bibitem[{\citenamefont{{Kravtsov} et~al.}(2006)\citenamefont{{Kravtsov},
  {Vikhlinin}, and {Nagai}}}]{Kravtsov06a}
\bibinfo{author}{\bibfnamefont{A.~V.} \bibnamefont{{Kravtsov}}},
  \bibinfo{author}{\bibfnamefont{A.}~\bibnamefont{{Vikhlinin}}},
  \bibnamefont{and} \bibinfo{author}{\bibfnamefont{D.}~\bibnamefont{{Nagai}}},
  \bibinfo{journal}{\apj} \textbf{\bibinfo{volume}{650}}, \bibinfo{pages}{128}
  (\bibinfo{year}{2006}), \eprint{arXiv:astro-ph/0603205}.

\bibitem[{\citenamefont{{Vikhlinin}
  et~al.}(2009{\natexlab{c}})\citenamefont{{Vikhlinin}, {Burenin}, {Ebeling},
  {Forman}, {Hornstrup}, {Jones}, {Kravtsov}, {Murray}, {Nagai}, {Quintana}
  et~al.}}]{Vikhlinin09d}
\bibinfo{author}{\bibfnamefont{A.}~\bibnamefont{{Vikhlinin}}},
  \bibinfo{author}{\bibfnamefont{R.~A.} \bibnamefont{{Burenin}}},
  \bibinfo{author}{\bibfnamefont{H.}~\bibnamefont{{Ebeling}}},
  \bibinfo{author}{\bibfnamefont{W.~R.} \bibnamefont{{Forman}}},
  \bibinfo{author}{\bibfnamefont{A.}~\bibnamefont{{Hornstrup}}},
  \bibinfo{author}{\bibfnamefont{C.}~\bibnamefont{{Jones}}},
  \bibinfo{author}{\bibfnamefont{A.~V.} \bibnamefont{{Kravtsov}}},
  \bibinfo{author}{\bibfnamefont{S.~S.} \bibnamefont{{Murray}}},
  \bibinfo{author}{\bibfnamefont{D.}~\bibnamefont{{Nagai}}},
  \bibinfo{author}{\bibfnamefont{H.}~\bibnamefont{{Quintana}}},
  \bibnamefont{et~al.}, \bibinfo{journal}{\apj} \textbf{\bibinfo{volume}{692}},
  \bibinfo{pages}{1033} (\bibinfo{year}{2009}{\natexlab{c}}),
  \eprint{arXiv:0805.2207}.

\bibitem[{\citenamefont{{Aghanim} et~al.}(2009)\citenamefont{{Aghanim}, {da
  Silva}, and {Nunes}}}]{Aghanim09a}
\bibinfo{author}{\bibfnamefont{N.}~\bibnamefont{{Aghanim}}},
  \bibinfo{author}{\bibfnamefont{A.~C.} \bibnamefont{{da Silva}}},
  \bibnamefont{and} \bibinfo{author}{\bibfnamefont{N.~J.}
  \bibnamefont{{Nunes}}}, \bibinfo{journal}{\aap}
  \textbf{\bibinfo{volume}{496}}, \bibinfo{pages}{637} (\bibinfo{year}{2009}),
  \eprint{arXiv:0808.0385}.

\bibitem[{\citenamefont{{Tinker} et~al.}(2008)\citenamefont{{Tinker},
  {Kravtsov}, {Klypin}, {Abazajian}, {Warren}, {Yepes}, {Gottl{\"o}ber}, and
  {Holz}}}]{Tinker08a}
\bibinfo{author}{\bibfnamefont{J.}~\bibnamefont{{Tinker}}},
  \bibinfo{author}{\bibfnamefont{A.~V.} \bibnamefont{{Kravtsov}}},
  \bibinfo{author}{\bibfnamefont{A.}~\bibnamefont{{Klypin}}},
  \bibinfo{author}{\bibfnamefont{K.}~\bibnamefont{{Abazajian}}},
  \bibinfo{author}{\bibfnamefont{M.}~\bibnamefont{{Warren}}},
  \bibinfo{author}{\bibfnamefont{G.}~\bibnamefont{{Yepes}}},
  \bibinfo{author}{\bibfnamefont{S.}~\bibnamefont{{Gottl{\"o}ber}}},
  \bibnamefont{and} \bibinfo{author}{\bibfnamefont{D.~E.}
  \bibnamefont{{Holz}}}, \bibinfo{journal}{\apj}
  \textbf{\bibinfo{volume}{688}}, \bibinfo{pages}{709} (\bibinfo{year}{2008}),
  \eprint{arXiv:0803.2706}.

\bibitem[{\citenamefont{{Ma}}(2007)}]{Ma07a}
\bibinfo{author}{\bibfnamefont{Z.}~\bibnamefont{{Ma}}}, \bibinfo{journal}{\apj}
  \textbf{\bibinfo{volume}{665}}, \bibinfo{pages}{887} (\bibinfo{year}{2007}),
  \eprint{arXiv:astro-ph/0610213}.

\bibitem[{\citenamefont{{Linder} and {White}}(2005)}]{Linder05b}
\bibinfo{author}{\bibfnamefont{E.~V.} \bibnamefont{{Linder}}} \bibnamefont{and}
  \bibinfo{author}{\bibfnamefont{M.}~\bibnamefont{{White}}},
  \bibinfo{journal}{\prd} \textbf{\bibinfo{volume}{72}},
  \bibinfo{pages}{061304} (\bibinfo{year}{2005}),
  \eprint{arXiv:astro-ph/0508401}.

\bibitem[{\citenamefont{{Bartelmann} et~al.}(2006)\citenamefont{{Bartelmann},
  {Doran}, and {Wetterich}}}]{Bartelmann06a}
\bibinfo{author}{\bibfnamefont{M.}~\bibnamefont{{Bartelmann}}},
  \bibinfo{author}{\bibfnamefont{M.}~\bibnamefont{{Doran}}}, \bibnamefont{and}
  \bibinfo{author}{\bibfnamefont{C.}~\bibnamefont{{Wetterich}}},
  \bibinfo{journal}{\aap} \textbf{\bibinfo{volume}{454}}, \bibinfo{pages}{27}
  (\bibinfo{year}{2006}), \eprint{arXiv:astro-ph/0507257}.

\bibitem[{\citenamefont{{Liberato} and {Rosenfeld}}(2006)}]{Liberato06a}
\bibinfo{author}{\bibfnamefont{L.}~\bibnamefont{{Liberato}}} \bibnamefont{and}
  \bibinfo{author}{\bibfnamefont{R.}~\bibnamefont{{Rosenfeld}}},
  \bibinfo{journal}{Journal of Cosmology and Astro-Particle Physics}
  \textbf{\bibinfo{volume}{7}}, \bibinfo{pages}{9} (\bibinfo{year}{2006}),
  \eprint{arXiv:astro-ph/0604071}.

\bibitem[{\citenamefont{{Basilakos} and {Voglis}}(2007)}]{Basilakos07a}
\bibinfo{author}{\bibfnamefont{S.}~\bibnamefont{{Basilakos}}} \bibnamefont{and}
  \bibinfo{author}{\bibfnamefont{N.}~\bibnamefont{{Voglis}}},
  \bibinfo{journal}{\mnras} \textbf{\bibinfo{volume}{374}},
  \bibinfo{pages}{269} (\bibinfo{year}{2007}), \eprint{arXiv:astro-ph/0610184}.

\bibitem[{\citenamefont{{Francis} et~al.}(2007)\citenamefont{{Francis},
  {Lewis}, and {Linder}}}]{Francis07a}
\bibinfo{author}{\bibfnamefont{M.~J.} \bibnamefont{{Francis}}},
  \bibinfo{author}{\bibfnamefont{G.~F.} \bibnamefont{{Lewis}}},
  \bibnamefont{and} \bibinfo{author}{\bibfnamefont{E.~V.}
  \bibnamefont{{Linder}}}, \bibinfo{journal}{\mnras}
  \textbf{\bibinfo{volume}{380}}, \bibinfo{pages}{1079} (\bibinfo{year}{2007}),
  \eprint{arXiv:0704.0312}.

\bibitem[{\citenamefont{{Francis}
  et~al.}(2009{\natexlab{a}})\citenamefont{{Francis}, {Lewis}, and
  {Linder}}}]{Francis09a}
\bibinfo{author}{\bibfnamefont{M.~J.} \bibnamefont{{Francis}}},
  \bibinfo{author}{\bibfnamefont{G.~F.} \bibnamefont{{Lewis}}},
  \bibnamefont{and} \bibinfo{author}{\bibfnamefont{E.~V.}
  \bibnamefont{{Linder}}}, \bibinfo{journal}{\mnras}
  \textbf{\bibinfo{volume}{393}}, \bibinfo{pages}{L31}
  (\bibinfo{year}{2009}{\natexlab{a}}), \eprint{arXiv:0810.0039}.

\bibitem[{\citenamefont{{Francis}
  et~al.}(2009{\natexlab{b}})\citenamefont{{Francis}, {Lewis}, and
  {Linder}}}]{Francis09b}
\bibinfo{author}{\bibfnamefont{M.~J.} \bibnamefont{{Francis}}},
  \bibinfo{author}{\bibfnamefont{G.~F.} \bibnamefont{{Lewis}}},
  \bibnamefont{and} \bibinfo{author}{\bibfnamefont{E.~V.}
  \bibnamefont{{Linder}}}, \bibinfo{journal}{\mnras}
  \textbf{\bibinfo{volume}{394}}, \bibinfo{pages}{605}
  (\bibinfo{year}{2009}{\natexlab{b}}), \eprint{arXiv:0808.2840}.

\bibitem[{\citenamefont{{Grossi} and {Springel}}(2009)}]{Grossi09a}
\bibinfo{author}{\bibfnamefont{M.}~\bibnamefont{{Grossi}}} \bibnamefont{and}
  \bibinfo{author}{\bibfnamefont{V.}~\bibnamefont{{Springel}}},
  \bibinfo{journal}{\mnras} \textbf{\bibinfo{volume}{394}},
  \bibinfo{pages}{1559} (\bibinfo{year}{2009}), \eprint{arXiv:0809.3404}.

\bibitem[{\citenamefont{{Fedeli} et~al.}(2009)\citenamefont{{Fedeli},
  {Moscardini}, and {Bartelmann}}}]{Fedeli09a}
\bibinfo{author}{\bibfnamefont{C.}~\bibnamefont{{Fedeli}}},
  \bibinfo{author}{\bibfnamefont{L.}~\bibnamefont{{Moscardini}}},
  \bibnamefont{and}
  \bibinfo{author}{\bibfnamefont{M.}~\bibnamefont{{Bartelmann}}},
  \bibinfo{journal}{\aap} \textbf{\bibinfo{volume}{500}}, \bibinfo{pages}{667}
  (\bibinfo{year}{2009}), \eprint{arXiv:0812.1097}.

\bibitem[{\citenamefont{{Casarini} et~al.}(2009)\citenamefont{{Casarini},
  {Macci{\`o}}, and {Bonometto}}}]{Casarini09a}
\bibinfo{author}{\bibfnamefont{L.}~\bibnamefont{{Casarini}}},
  \bibinfo{author}{\bibfnamefont{A.~V.} \bibnamefont{{Macci{\`o}}}},
  \bibnamefont{and} \bibinfo{author}{\bibfnamefont{S.~A.}
  \bibnamefont{{Bonometto}}}, \bibinfo{journal}{Journal of Cosmology and
  Astro-Particle Physics} \textbf{\bibinfo{volume}{3}}, \bibinfo{pages}{14}
  (\bibinfo{year}{2009}), \eprint{arXiv:0810.0190}.

\bibitem[{\citenamefont{{Knox} et~al.}(2006)\citenamefont{{Knox}, {Song}, and
  {Zhan}}}]{Knox06b}
\bibinfo{author}{\bibfnamefont{L.}~\bibnamefont{{Knox}}},
  \bibinfo{author}{\bibfnamefont{Y.-S.} \bibnamefont{{Song}}},
  \bibnamefont{and} \bibinfo{author}{\bibfnamefont{H.}~\bibnamefont{{Zhan}}},
  \bibinfo{journal}{\apj} \textbf{\bibinfo{volume}{652}}, \bibinfo{pages}{857}
  (\bibinfo{year}{2006}), \eprint{arXiv:astro-ph/0605536}.

\bibitem[{\citenamefont{{Hill} et~al.}(2008)\citenamefont{{Hill}, {Gebhardt},
  {Komatsu}, {Drory}, {MacQueen}, {Adams}, {Blanc}, {Koehler}, {Rafal}, {Roth}
  et~al.}}]{Hill08a}
\bibinfo{author}{\bibfnamefont{G.~J.} \bibnamefont{{Hill}}},
  \bibinfo{author}{\bibfnamefont{K.}~\bibnamefont{{Gebhardt}}},
  \bibinfo{author}{\bibfnamefont{E.}~\bibnamefont{{Komatsu}}},
  \bibinfo{author}{\bibfnamefont{N.}~\bibnamefont{{Drory}}},
  \bibinfo{author}{\bibfnamefont{P.~J.} \bibnamefont{{MacQueen}}},
  \bibinfo{author}{\bibfnamefont{J.}~\bibnamefont{{Adams}}},
  \bibinfo{author}{\bibfnamefont{G.~A.} \bibnamefont{{Blanc}}},
  \bibinfo{author}{\bibfnamefont{R.}~\bibnamefont{{Koehler}}},
  \bibinfo{author}{\bibfnamefont{M.}~\bibnamefont{{Rafal}}},
  \bibinfo{author}{\bibfnamefont{M.~M.} \bibnamefont{{Roth}}},
  \bibnamefont{et~al.}, in \emph{\bibinfo{booktitle}{Astronomical Society of
  the Pacific Conference Series}}, edited by
  \bibinfo{editor}{\bibfnamefont{T.}~\bibnamefont{{Kodama}}},
  \bibinfo{editor}{\bibfnamefont{T.}~\bibnamefont{{Yamada}}}, \bibnamefont{and}
  \bibinfo{editor}{\bibfnamefont{K.}~\bibnamefont{{Aoki}}}
  (\bibinfo{year}{2008}), vol. \bibinfo{volume}{399} of
  \emph{\bibinfo{series}{Astronomical Society of the Pacific Conference
  Series}}, p. \bibinfo{pages}{115}, \eprint{arXiv:0806.0183}.

\bibitem[{\citenamefont{{Zhan} et~al.}(2009)\citenamefont{{Zhan}, {Knox}, and
  {Tyson}}}]{Zhan09a}
\bibinfo{author}{\bibfnamefont{H.}~\bibnamefont{{Zhan}}},
  \bibinfo{author}{\bibfnamefont{L.}~\bibnamefont{{Knox}}}, \bibnamefont{and}
  \bibinfo{author}{\bibfnamefont{J.~A.} \bibnamefont{{Tyson}}},
  \bibinfo{journal}{\apj} \textbf{\bibinfo{volume}{690}}, \bibinfo{pages}{923}
  (\bibinfo{year}{2009}), \eprint{arXiv:0806.0937}.

\end{thebibliography}

\end{document}